\documentclass[lettersize,journal]{IEEEtran}
\usepackage{amsmath,amsfonts}
\usepackage{algorithmic}
\usepackage{algorithm}
\usepackage{array}
\usepackage{caption}
\usepackage[caption=false,font=normalsize,labelfont=sf,textfont=sf]{subfig}
\usepackage{textcomp}
\usepackage{stfloats}
\usepackage{url}
\usepackage{verbatim}
\usepackage{graphicx}
\usepackage{epsfig}
\usepackage{epstopdf}
\newenvironment{proof}{{\indent \indent \it Proof:}}{\hfill $\blacksquare$\par}

\usepackage{amssymb}
\usepackage{cite}
\usepackage{booktabs}
\newtheorem{proposition}{\textbf{Proposition}}
\allowdisplaybreaks[1]
\usepackage{balance}

\def\BibTeX{{\rm B\kern-.05em{\sc i\kern-.025em b}\kern-.08em
    T\kern-.1667em\lower.7ex\hbox{E}\kern-.125emX}}
\begin{document}
\title{Channel Estimation for RIS-Aided MU-MIMO mmWave Systems with Practical Hybrid Architecture}
\author{Liuchang Zhuo, Cunhua Pan,~\IEEEmembership{Senior Member,~IEEE}, Hong Ren,~\IEEEmembership{Member,~IEEE}, Ruisong Weng, Shi Jin,~\IEEEmembership{Fellow,~IEEE}, A. Lee Swindlehurst,~\IEEEmembership{Fellow,~IEEE}, and Jiangzhou Wang,~\IEEEmembership{Fellow,~IEEE}
	
\thanks{Liuchang Zhuo, Cunhua Pan, Hong Ren, Ruisong Weng, Shi Jin and Jiangzhou Wang are with National Mobile Communications Research Laboratory, Southeast University, Nanjing, China (e-mail:{230248854, cpan, hren, ruisong\_weng, jinshi, j.z.wang}@seu.edu.cn). 
	
A. Lee Swindlehurst is with the Center for Pervasive Communications
and Computing, University of California, Irvine, CA 92697 USA (e-mail:
swindle@uci.edu).}
}

\maketitle
\begin{abstract}
This paper proposes a correlation-based three-stage channel estimation strategy with low pilot overhead for reconfigurable intelligent surface (RIS)-aided millimeter wave (mmWave) multi-user (MU) MIMO systems, in which both users and base station (BS) are equipped with a hybrid RF architecture. In Stage I, all users jointly transmit pilots and recover the uncompressed received signals to estimate the angle of arrival (AoA) at the BS using the discrete Fourier transform (DFT). Based on the observation that the overall cascaded MIMO channel can be decomposed into multiple sub-channels, the cascaded channel for a typical user is estimated in Stage II. Specifically, using the invariance of angles and the linear correlation of gains related to different cascaded subchannels, we use compressive sensing (CS), 
least squares (LS), and a one-dimensional search to estimate the Angles of Departure (AoDs), based on which the overall cascaded channel is obtained. In Stage III, the remaining users independently transmit pilots to estimate their individual cascaded channel with the same approach as in Stage II, which exploits the equivalent common RIS-BS channel obtained in Stage II to reduce the pilot overhead. In addition, the hybrid combining matrix and the RIS phase shift matrix are designed to reduce the noise power, thereby further improving the estimation performance. Simulation results demonstrate that the proposed algorithm can achieve high estimation accuracy especially when the number of antennas at the users is small, and reduce pilot overhead by more than five times compared with the existing benchmark approach.
\end{abstract}


\section{INTRODUCTION}
As one of the core technologies of fifth generation (5G) wireless communication systems, millimeter wave (mmWave) communications is intended to provide high data rates and low latency capabilities, supporting the development of the automation industry. However, mmWave signals are subject to blockages, which leads to limited coverage ranges and blind spots\cite{alliance2023reconfigurable}. Fortunately, reconfigurable intelligent surface (RIS) technology is a promising tool that can effectively compensate for such shortcomings in mmWave systems \cite{9475160,8910627,9140329,you2021towards,9110849}. An RIS is usually composed of a large number of carefully designed adjustable electromagnetic units whose properties can be dynamically controlled \cite{cui2014coding}. This allows an RIS to actively enrich the channel scattering conditions and obtain additional multiplexing gain. The prerequisite for fully reaping this performance gain is to acquire accurate channel state information (CSI) \cite{9847080,9771077,liang2021reconfigurable,zheng2022survey}. However, due to its passive nature, a purely passive RIS is unable to receive and process signals, which makes it difficult to separately estimate the RIS-BS and user-RIS channels. 

Fortunately, the feasibility of transmission design based only on the cascaded channels has been demonstrated \cite{9076830,9014322}. As a result, most existing contributions have focused on estimating the cascaded channel for RIS-aided mmWave systems \cite{9103231,9328485,10053657,9732214,9760391,9919846}. For example, to reduce pilot overhead, \cite{9103231} exploits the spatial sparsity of mmWave channels and proposes a channel estimation method based on compressive sensing (CS) for a single-user (SU) system. For the MU-MISO case, \cite{9328485,10053657} develop a method based on the double sparse structure of the cascaded channel and the common RIS-BS channel. However, the pilot overhead in \cite{10053657} increases at an unacceptable rate as the number of RIS elements grows, which greatly limits the performance gain. To further reduce the pilot overhead, \cite{9732214} derives the angle-gain scaling property between the cascaded multipaths and is able to reduce the pilot overhead to a low level. The authors of \cite{9760391} propose a CS-based estimation (CS-EST) algorithm for more general MIMO systems, but it cannot be efficiently used in the multi-user case since it does not exploit the common RIS-BS channel shared by all users. However the authors of \cite{9919846} extend the estimation strategy of \cite{9732214} to the MU-MIMO case.

All the above-mentioned channel estimation schemes consider fully-digital architectures, which are costly to implement in mmWave communication systems with large antenna arrays. An alternative hybrid architecture can be employed in which the number of radio frequency (RF) chains is much lower than the total number of antennas since each RF chain is connected to multiple active antennas via a network of analog phase shifters \cite{6736750}. The authors of \cite{7389996} demonstrated that hybrid beamforming can approach the same performance of the fully-digital architecture with many fewer RF chains, on the order of only twice the number of transmitted data streams. For the hybrid case, the dimension of the received signal is much lower than that of the antenna array, which complicates the recovery of CSI. One possible approach is to directly extend existing estimation methods from the fully-digital case, i.e. \cite{9760391} and \cite{9919846}, to hybrid systems based on \cite{8493600}. However, this approach has two shortcomings: First, due to the necessity of restoring the received signals before compression, when the number of antennas is much larger than the number of RF chains, the pilot overhead will be excessive. Second, it is difficult for existing methods to estimate the AoDs at user equipment (UE) with a small number of antennas. For example, the simultaneous orthogonal matching pursuit (SOMP) algorithm is used in both \cite{9760391} and \cite{9919846} to estimate the UE AoDs, but when the number of UE antennas is small, the strong correlation between atoms in the overcomplete dictionary can lead to poor performance.
To solve this problem, \cite{9398559} proposes a two-stage channel estimation strategy based on atomic norm minimization (ANM) for hybrid 
structures. However, this approach has high complexity and requires excessive pilot overhead in multiple-UE cases. 

Based on the above discussion, we propose a correlation-based three-stage channel estimation strategy for RIS-aided mmWave MU-MIMO systems, in which both the BS and UEs are equipped with hybrid RF architectures. The main contributions of this work are summarized as follows:

\begin{itemize}
	\item We propose a novel form of the cascaded channel decomposition for an RIS-aided MU-MIMO communication system in which the cascaded MIMO channel is decomposed into several MISO subchannels. Based on this decomposition, we propose a three-stage estimation framework that is suitable for hybrid architectures. In particular, in Stage I, the common AoAs at the BS are estimated through a DFT-based approach. Stage II has two sub-stages, and only a typical user sends pilots to the BS for channel estimation. In the first sub-stage, the typical user uses the $1$-st antenna to send pilots to estimate the $1$-st subchannel and construct the equivalent common RIS-BS channel, which is used to reduce the pilot overhead in Stage III. In the second sub-stage, only a small number of the available antennas is used to send pilots to estimate the remaining subchannels, which are then combined into the overall cascaded channel. Similar to Stage II, Stage III estimates the cascaded channels of all the remaining users via two sub-stages with low pilot overhead using the re-parameterized RIS-BS channel.
	
	\item We demonstrate that the proposed method greatly reduces the pilot overhead. Unlike \cite{9919846}, rather than using \cite{8493600} to extend the approach to hybrid systems, we divide Stage I in \cite{9919846} into two stages. The common AoAs are estimated in Stage I, and the cascaded channel of a typical user is estimated in Stage II. Under the assumption that the number of RF chains at the BS is larger than the number of propagation paths from the RIS to the BS, we use the method of \cite{8493600} to recover only the uncompressed received signals in Stage I to guarantee the performance of the DFT-based method, while the received signals are directly used to perform channel estimation in Stage II and Stage III, an approach that significantly reduces the pilot overhead.

	\item We improve the channel estimation accuracy when the number of UE antennas is small by using the inherent relationship between different 
    subchannels. Specifically, when using different antennas to send pilot signals, the cascaded cosine angles remain unchanged while the scaling factors of the cascaded gains are related to the AoDs. Based on this property, in Stage II and Stage III we divide estimation of the overall cascaded channel into two parts. In the first part, CS is exploited to estimate the cascaded cosine angles and gains related to the first 
    subchannel. In the second, we exploit the invariance of the cascaded cosine angles to estimate the cascaded gains of other subchannels using least squares (LS) with low computational complexity. To further reduce pilot overhead, only a small fraction of the cascaded gains are estimated using LS, and the results can be used to determine the UE AoDs using a one-dimensional search with low complexity. Then, all  remaining subchannels can be found based on the AoD steering matrix. Because this one-dimensional search is performed using a super-resolution approach, it effectively avoids the strict requirements on the number of UE antennas needed to ensure high estimation accuracy when using the SOMP algorithm in \cite{9760391,9919846}.
	
\end{itemize}

The rest of the paper is organized as follows. Section II introduces the hybrid system model and a novel decomposition method for the cascaded channel. The proposed three-stage channel estimation protocol is investigated in Section III. Both the hybrid combining matrices and the RIS phase shift matrices are optimized in Section IV. Section V provides simulation results, followed by conclusions in Section VI.

Notation: The following mathematical notation and symbols are used throughout the paper. The imaginary unit is defined as $i\triangleq \sqrt{-1}$. Scalars and sets are represented by lowercase and calligraphy letters, respectively, i.e., $a$ is a scalar and $\mathcal{A} $ is a set. The operation $[a]$ rounds up to the nearest integer. Matrices and vectors are denoted by boldface uppercase and lowercase letters, respectively, i.e., $\mathbf{a}$ is a vector and $\mathbf{A}$ is a matrix. Statistical expectation is denoted by $\mathbb{E} \{ \cdot \}$, and $\mathbf{A}^{\mathrm{T}}$, $\mathbf{A}^*$, $\mathbf{A}^{\mathrm{H}}$, $\mathbf{A}^{\dagger}$, and $\| \mathbf{A} \| _F$ are the transpose, conjugate, conjugate transpose, left pseudo-inverse, and Frobenius norm of matrix $\mathbf{A}$, respectively. The $m$-th element of vector $\mathbf{a}$ is given by $[\mathbf{a}]_m$ and the $(m,n)$-th element of matrix $\mathbf{A}$ is denoted by $[ \mathbf{A} ] _{m,n}$. The $m$-th row and $n$-th column of matrix $\mathbf{A}$ are represented using $\left[ \mathbf{A} \right] _{m,:}$ and $\left[ \mathbf{A} \right] _{:,n}$, respectively. A diagonal matrix with the entries of vector $\mathbf{a}$ on its main diagonal is written as $\mathrm{diag}\left\{ \mathbf{a} \right\}$. The Khatri-Rao product between two matrices $\mathbf{A}$ and $\mathbf{B}$ is denoted by $\mathbf{A}\diamond \mathbf{B}$.

\section{SYSTEM MODEL}
We consider a narrow-band time-division duplex (TDD) mmWave massive MIMO system where a BS communicates with $K$ users. We assume that the BS is equipped with an $N_{\mathrm{bs}}$-antenna uniform linear array (ULA) and $N_{\mathrm{rf}}$ RF chains, and the $k$-th user also employs a ULA with $Q_k$ antennas and $Q_{\mathrm{rf},k}$ RF chains. To improve communication performance, an RIS equipped with a uniform planar array (UPA) of dimension $M=M_1\times M_2$ is deployed between the BS and the UEs. Based on the assumption of channel reciprocity, the CSI of the downlink channel is acquired by estimating the uplink channel. We assume that the direct channel between the BS and UEs is blocked. 

We respectively denote the AoA of the $l$-th path from the RIS to the BS and the AoD of the $j_k$-th path from user $k$ to the RIS as $\vartheta_l$ and $\zeta _{k,j_k}$. The array response vector of a ULA with $N$ elements can be represented as
\setlength{\abovedisplayskip}{3pt}
\setlength{\belowdisplayskip}{3pt}
\begin{align}\label{ULA_array}
	\mathbf{a}_X( x )=[ 1,e^{-j2\pi x},\cdots ,e^{-j2\pi ( N-1 ) x} ] ^{\mathrm{T}},
\end{align}
where $X\in \{ N_{\mathrm{bs}},Q_k \}$, $x\in \{ \psi _l,\xi _{k,j_k} \} $, $\psi _l=\frac{d_{\mathrm{bs}}}{\lambda}\cos ( \vartheta _l )$ and $\xi _{k,j_k}=\frac{d_{\mathrm{ue}}}{\lambda}\cos ( \zeta _{k,j_k} )$ are the directional cosines, $d_{\mathrm{bs}}$ and $d_{\mathrm{ue}}$ are the element spacings of the BS and UE antennas, respectively, and $\lambda$ is the carrier wavelength. Similarly, we respectively denote $\nu _l$ and $\varpi _l$ as the elevation and azimuth AoDs of the $l$-th path from the RIS to the BS, and $\epsilon _{k,j_k}$ and $\phi _{k,j_k}$ as the elevation and azimuth AoAs of the $j_k$-th spatial path from user $k$ to the RIS. The steering vector of an $M=M_1\times M_2$ UPA can be represented as
\begin{align}
	\mathbf{a}_M( y,z ) =\mathbf{a}_{M_1}( y ) \otimes \mathbf{a}_{M_2}( z ),
\end{align}
where $y\in \{ \upsilon _l,\theta _{k,j_k} \} $, $z\in \{ \omega _l,\varphi _{k,j_k} \} $, and $\mathbf{a}_{M_1}( y )$ and $\mathbf{a}_{M_2}( z )$ are the steering vectors with respect to the vertical and horizontal directions, respectively. Let the spacing between the RIS elements be $d_{\mathrm{ris}}$, so that the relationships between the spatial frequencies and their corresponding angles are given by $\upsilon _l=\frac{d_{\mathrm{ris}}}{\lambda}\cos ( \nu _l )$, $\omega _l=\frac{d_{\mathrm{ris}}}{\lambda}\cos ( \varpi _l ) \sin ( \nu _l ) $, $\theta _{k,j_k}=\frac{d_{\mathrm{ris}}}{\lambda}\cos ( \epsilon _{k,j_k} )$ and $\varphi _{k,j_k}=\frac{d_{\mathrm{ris}}}{\lambda}\cos( \phi _{k,j_k})\sin ( \epsilon _{k,j_k})$. 
Using the geometric parameters defined above, the channel from the RIS to the BS, denoted by $\mathbf{H}_{\mathrm{br}}\in \mathbb{C} ^{N_{\mathrm{bs}}\times M}$, and the channel between user $k$ and the RIS, denoted by $\mathbf{H}_k\in \mathbb{C} ^{M\times Q_k}$, can be represented as
\begin{subequations}\label{geometric_channel_model}
	\begin{align}
		\mathbf{H}_{\mathrm{br}}=&\sum_{l=1}^L{\alpha _l\mathbf{a}_{N_{\mathrm{bs}}}(\psi _l) \mathbf{a}_{M}^{\mathrm{H}}( \upsilon _l,\omega _l )},\label{H_br}
		\\
		\mathbf{H}_k=&\sum_{j_k=1}^{J_k}{\beta _{j_k}\mathbf{a}_M(\theta _{k,j_k},\varphi _{k,j_k})\mathbf{a}_{Q_k}^{\mathrm{H}}(\xi _{k,j_k})},\label{H_k}
	\end{align}
\end{subequations}
where $L$ and $J_k$ denote the number of propagation paths from the RIS to the BS and between user $k$ and the RIS, respectively, the complex path gains of the $l$-th path in the RIS-BS channel and that of the $j_k$-th path in the user $k$-RIS channel are respectively represented as $\alpha _l$ and $\beta _{j_k}$. 

For the RIS-BS channel, the AoA array response matrix, complex gain matrix and AoD array response matrix are denoted as $\mathbf{A}_{N_{\mathrm{bs}}}=\left[ \mathbf{a}_{N_{\mathrm{bs}}}\left( \psi _1 \right) ,\cdots ,\mathbf{a}_{N_{\mathrm{bs}}}\left( \psi _L \right) \right]\in \mathbb{C} ^{N_{\mathrm{bs}}\times L}$, $\mathbf{\Lambda }=\mathrm{diag}\left\{ \alpha _1,\cdots ,\alpha _L \right\}=\mathrm{diag}\{\boldsymbol{\alpha}\}\in \mathbb{C} ^{L\times L}$ and $\mathbf{A}_M=\left[ \mathbf{a}_M\left( \upsilon _1,\omega _1 \right) ,\cdots ,\mathbf{a}_M\left( \upsilon _L,\omega _L \right) \right]\in \mathbb{C} ^{M\times L}$, respectively. For the user $k$-RIS channel, the AoA array response matrix, complex gain matrix and AoD array response matrix are defined as $\mathbf{A}_{M,k}=\left[ \mathbf{a}_M( \theta _{k,1},\varphi _{k,1} ) ,\cdots ,\mathbf{a}_M( \theta _{k,J_k},\varphi _{k,J_k} ) \right]\in \mathbb{C} ^{M\times J_k}$, $\mathbf{B}_k=\mathrm{diag}\left\{ \beta _{k,1},\cdots ,\beta _{k,J_k} \right\}=\mathrm{diag}\{\boldsymbol{\beta _k}\}\in \mathbb{C} ^{J_k\times J_k}$ and $\mathbf{A}_{Q_k}=\left[ \mathbf{a}_{Q_k}(\xi _{k,1}),\cdots ,\mathbf{a}_{Q_k}(\xi _{k,J_k}) \right]\in \mathbb{C} ^{Q_k\times J_k}$, respectively. Thus, $\mathbf{H}_{\mathrm{br}}$ and $\mathbf{H}_k$ in \eqref{geometric_channel_model} can be written using the following compact notation: 
\begin{subequations}\label{MIMO_channel_model}
	\begin{align}
		\mathbf{H}_{\mathrm{br}}=&\mathbf{A}_{N_{\mathrm{bs}}}\mathbf{\Lambda A}_{M}^{\mathrm{H}},\label{H_matrix}
		\\
		\mathbf{H}_{k}=&\mathbf{A}_{M,k}\mathbf{B}_k\mathbf{A}_{Q_k}^{\mathrm{H}}.\label{H_k_matrix}
	\end{align}
\end{subequations}

Since an RIS is unable to demodulate and process its received signals, it is generally not possible to separately estimate the user-RIS channel and the RIS-BS channel. Thus our work focuses on estimation of the cascaded channels
\begin{align}\label{MIMO_G_k_easy}
	\mathbf{G}_k=\mathbf{H}_{k}^{\mathrm{T}}\diamond \mathbf{H}_{\mathrm{br}}, 1\le k\le K.
\end{align}
Defining $\mathbf{h}_{k,q_k}=[ \mathbf{H}_k ] _{:,q_k}$, and using the decomposition of the Khatri-Rao product between two matrices in \cite{10137372}, the subchannel for UE $k$ corresponding to its $q_k$-th antenna can be written as
\begin{align}\label{MIMO_G_k_qk}
	\mathbf{G}_{k,q_k}=\mathbf{H}_{\mathrm{br}}\mathrm{diag}\{ [ \mathbf{H}_k ] _{:,q_k} \}= \mathbf{H}_{\mathrm{br}}\mathrm{diag}\{ \mathbf{h}_{k,q_k} \}.
\end{align}
Combining \eqref{MIMO_G_k_easy} with \eqref{MIMO_G_k_qk}, $\mathbf{G}_k$ can be rewritten as a combination of $Q_k$ subchannels: 
\begin{align}\label{MIMO_G_k_complex}
	{\mathbf{G}}_k&=\mathbf{H}_{k}^{\mathrm{T}}\diamond \mathbf{H}_{\mathrm{br}}\nonumber
	\\
	&=[ ( \mathbf{H}_{\mathrm{br}}\mathrm{diag}\{ [ \mathbf{H}_{k} ] _{:,1} \} ) ^{\mathrm{T}},\cdots ,( \mathbf{H}_{\mathrm{br}}\mathrm{diag}\{ [ \mathbf{H}_{k} ] _{:,Q_k} \} ) ^{\mathrm{T}} ] ^{\mathrm{T}}\nonumber
	\\
	&=[ \mathbf{G}_{k,1}^{\mathrm{T}},\cdots ,\mathbf{G}_{k,Q_k}^{\mathrm{T}} ] ^{\mathrm{T}}.
\end{align}

Define $\mathbf{W}_{i}^{\left( j \right)}\in \mathbb{C} ^{N_{\mathrm{rf}}\times N_{\mathrm{bs}}}$ and $\mathbf{e}_{i}^{\left( j \right)}\in \mathbb{C} ^{M\times 1}$ as the hybrid combining matrix at the BS and the vector containing the RIS phase shifts at the $j$-th time slot of the $i$-th frame, respectively. Further define the hybrid precoding matrix of UE $k$ as $\mathbf{F}_{k,i}^{\left( j \right)}\in \mathbb{C} ^{Q_{k}\times Q_{\mathrm{rf},k}}$. Then, during the uplink transmission in the $j$-th time slot of the $i$-th frame, the received signals can be expressed as
\begin{align}\label{signal_model_MIMO}
	\mathbf{y}_{k,i}^{\left( j \right)}=\mathbf{W}_{i}^{\left( j \right)}\mathbf{H}_{\mathrm{br}}\mathrm{diag}\{\mathbf{e}_{i}^{\left( j \right)}\}\mathbf{H}_k\mathbf{F}_{k,i}^{\left( j \right)}\mathbf{s}_{k,i}^{\left( j \right)}+\mathbf{W}_{i}^{\left( j \right)}\mathbf{n}_{i}^{\left( j \right)},
\end{align}
where $\mathbf{s}_{k,i}^{( j )}\in \mathbb{C} ^{Q_{\mathrm{rf},k}\times 1}$ are the pilot signals, which satisfy $| [ \mathbf{s}_{k,i}^{( j )} ] _m |=1$ for $1 \le m \le Q_{\mathrm{rf},k}$, $P_{k}$ is the total transmission power of user $k$, which satisfies $P_k=\| \mathbf{F}_{k,i}^{\left( j \right)}\mathbf{s}_{k,i}^{\left( j \right)} \| _{2}^{2}$, and $\mathbf{n}_{i}^{\left( j \right)}\in \mathbb{C} ^{N_{\mathrm{bs}}\times 1}\sim \mathcal{C} \mathcal{N} ( \mathbf{0},\sigma ^2\mathbf{I}_{N_{\mathrm{bs}}} )$ represents additive white Gaussian noise (AWGN) with power $\sigma ^2$ at the BS.

\section{Three-Stage Channel Estimation Method with Hybrid Architecture}
\subsection{Channel Estimation Protocol}
In this section, we propose a novel three-stage correlation-based uplink channel estimation strategy for the considered hybrid communication system. In Stage I, to fully exploit the MU diversity gain, all users simultaneously send pilot signals to estimate the common AoA matrix at the BS. We first recover the uncompressed received signals \cite{8493600} and then derive the AoAs at the BS using a DFT. In Stage II, the cascaded channel of a typical user (denoted as user 1) and the equivalent common RIS-BS channel are estimated. To reduce the pilot overhead, instead of recovering the uncompressed signals first as in Stage I, we directly use the compressed received signal for channel estimation. There are two sub-stages in Stage II. In particular, in the first sub-stage, user 1 only sends pilots from its first antenna to estimate just the first subchannel via OMP. Then, we construct an equivalent common RIS-BS channel that is used to reduce the pilot overhead in the subsequent estimation step. In the second sub-stage, instead of estimating all the remaining subchannels by sequentially transmitting pilots from the corresponding antennas, we only estimate a subset of the subchannels and derive the AoDs of UE 1 using a simple one-dimensional search. Based on the estimated AoD matrix, all the remaining 
subchannels can be obtained without sending additional pilots, and these estimated subchannels can be combined into the overall cascaded channel of UE $1$ using the relationship among different subchannels. In Stage III, the cascaded channels of all the remaining users are estimated. Similar to Stage II, the $1$-st subchannel is estimated in Sub-Stage I and all the remaining subchannels are found in Sub-Stage II. By exploiting the re-parameterized RIS-BS channel constructed in Stage II, fewer pilots are required in Stage III. 

\subsection{Stage I: Estimation of the common AoA matrix at the BS}
\begin{figure}[t]
	\centering
	\includegraphics[width=0.48\textwidth]{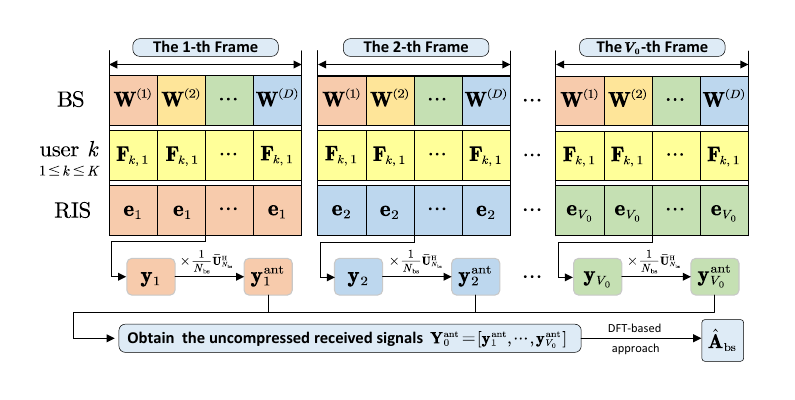}
	\caption{The working diagram of Stage I in the proposed channel estimation protocol.}
	\label{fig_StageI}
\end{figure}

In this stage, all the $K$ users simultaneously send pilot signals. As shown in Fig. \ref{fig_StageI}, $V_0$ frames are needed, where each frame has $D=\frac{N_{\mathrm{bs}}}{N_{\mathrm{rf}}}$ time slots. In each time slot, the phase shift vector of the RIS remains unchanged, and all the $K$ users use their $1$-st antenna to send a pilot signal, i.e., for $\forall d\in [ 1,D ]$ and $\forall i\in \left[ 1,V_0 \right]$, $\mathbf{e}_{i}^{\left( d \right)}=\mathbf{e}_{i}$ and $\mathbf{F}_{k,i}^{\left( d \right)}=\mathbf{F}_{k,1}=\frac{\sqrt{P_k}}{Q_{\mathrm{rf},k}}[\mathbf{1}_{Q_{\mathrm{rf},k}\times 1},\mathbf{0}_{Q_{\mathrm{rf},k}\times \left( Q_k-1 \right)}]^{\mathrm{T}}$. We define $\bar{\mathbf{U}}_{N_{\mathrm{bs}}}\in \mathbb{C} ^{N_{\mathrm{bs}}\times N_{\mathrm{bs}}}$ as the discrete DFT matrix whose $\left( n,m \right)$-th entry is given by $\left[ \bar{\mathbf{U}}_{N_{\mathrm{bs}}} \right] _{n,m}=e^{-j\frac{2\pi \left( n-1 \right) \left( m-1 \right)}{N_{\mathrm{bs}}}}$. Similar to \cite{8493600}, the hybrid combining matrix in the $d$-th time slot of the $i$-th frame is designed as
\begin{align}
	&\mathbf{W}_{i}^{\left( d \right)}=\mathbf{W}^{\left( d \right)}=\left[ \bar{\mathbf{U}}_{N_{\mathrm{bs}}} \right] _{\left( d-1 \right) N_{\mathrm{rf}}+1:dN_{\mathrm{rf}},:}\nonumber
	\\
	&=\left[ \begin{matrix}
		1&		e^{-j\frac{2\pi \left( d-1 \right) N_{\mathrm{rf}}}{N_{\mathrm{bs}}}}&		\cdots&		e^{-j\frac{2\pi \left( d-1 \right) N_{\mathrm{rf}}\left( N_{\mathrm{bs}}-1 \right)}{N_{\mathrm{bs}}}}\\
		\vdots&		\vdots&		\vdots&		\vdots\\
		1&		e^{-j\frac{2\pi \left( dN_{\mathrm{rf}}-1 \right)}{N_{\mathrm{bs}}}}&		\cdots&		e^{-j\frac{2\pi \left( dN_{\mathrm{rf}}-1 \right) \left( N_{\mathrm{bs}}-1 \right)}{N_{\mathrm{bs}}}}\\
	\end{matrix} \right].
\end{align}

For simplicity, the pilots sent by the $k$-th user are assumed to be a vector of all ones: $\mathbf{s}_{k,i}^{\left( d \right)}=\mathbf{1}_{Q_{\mathrm{rf},k}\times 1}$. The received signal in the $d$-th time slot of the $i$-th frame can be written as
\begin{align}\label{stage1_y}
	\mathbf{y}_{i}^{\left( d \right)}&=\mathbf{W}_{i}^{\left( d \right)}\sum_{k=1}^K{\mathbf{H}_{\mathrm{br}}\mathrm{diag}\{\mathbf{e}_{i}^{\left( d \right)}\}\mathbf{H}_k\mathbf{F}_{k,i}^{\left( d \right)}\mathbf{s}_{k,i}^{\left( d \right)}}+\mathbf{W}_{i}^{\left( d \right)}\mathbf{n}_{i}^{\left( d \right)}\nonumber
	\\
	&=\mathbf{W}^{\left( d \right)}\sum_{k=1}^K{\sqrt{P_k}\mathbf{H}_{\mathrm{br}}\mathrm{diag}\{\mathbf{e}_{i}\}[ \mathbf{H}_k ] _{:,1}}+\mathbf{W}_{i}^{( d )}\mathbf{n}_{i}^{( d )}\nonumber
	\\
	&=\mathbf{W}^{\left( d \right)}\sum_{k=1}^K{\sqrt{P_k}\mathbf{H}_{\mathrm{br}}\mathrm{diag}\{\mathbf{h}_{k,1}\}\mathbf{e}_{i}}+\mathbf{W}_{i}^{( d )}\mathbf{n}_{i}^{( d )},
\end{align} 
where $\mathbf{n}_{i}^{\left( d \right)}\sim \mathcal{C} \mathcal{N} \left( \mathbf{0},\sigma ^2\mathbf{I}_{N_{\mathrm{bs}}} \right)$ and $\mathbf{h}_{k,1}$ is defined in Section II as $\mathbf{h}_{k,1}=[ \mathbf{H}_k ] _{:,1}$. By stacking the $D$ signals received during the $i$-th frame, we have
\begin{align}\label{y_0_i_ant}
	\mathbf{y}_i&=[(\mathbf{y}_{i}^{\left( 1 \right)})^{\mathrm{T}},\cdots ,(\mathbf{y}_{i}^{\left( D \right)})^{\mathrm{T}}]^{\mathrm{T}}\nonumber
	\\
	&=\mathbf{W}_i\mathbf{H}_{\mathrm{br}}\sum_{k=1}^K{\sqrt{P_k}\mathrm{diag}\{\mathbf{h}_{k,1}\}}\mathbf{e}_i+\mathbf{n}_i,
\end{align}
where $\mathbf{W}_i=[(\mathbf{W}_{i}^{\left( 1 \right)})^{\mathrm{T}},\cdots ,(\mathbf{W}_{i}^{\left( D \right)})^{\mathrm{T}}]^{\mathrm{T}}=\bar{\mathbf{U}}_{N_{\mathrm{bs}}}$, and $\mathbf{n}_i=[(\mathbf{n}_{i}^{\left( 1 \right)})^{\mathrm{T}},\cdots ,(\mathbf{n}_{i}^{\left( D \right)})^{\mathrm{T}}]^{\mathrm{T}}$ is the corresponding noise term. Using the property $\bar{\mathbf{U}}_{N_{\mathrm{bs}}}\bar{\mathbf{U}}_{N_{\mathrm{bs}}}^{\mathrm{H}}=N_{\mathrm{bs}}$, an uncompressed received signal can be obtained in the $i$-th frame as follows:
\begin{align}
	\mathbf{y}_{i}^{\mathrm{ant}}&=\frac{1}{N_{\mathrm{bs}}}\bar{\mathbf{U}}_{N_{\mathrm{bs}}}^{\mathrm{H}}\mathbf{y}_i\nonumber
	\\
	&=\mathbf{H}_{\mathrm{br}}\sum_{k=1}^K{\sqrt{P_k}\mathrm{diag}\{\mathbf{h}_{k,1}\}}\mathbf{e}_i+\mathbf{n}_{i}^{\mathrm{ant}},
\end{align}
where $\frac{1}{N_{\mathrm{bs}}}\bar{\mathbf{U}}_{N_{\mathrm{bs}}}^{\mathrm{H}}\mathbf{n}_{i}^{\mathrm{ant}}$. By further stacking all the $V_0$ uncompressed received signals $\mathbf{y}_{i}^{\mathrm{ant}}$ in Stage I, we obtain
\begin{align}\label{Y_0_ant_MIMO}
	\mathbf{Y}_{0}^{\mathrm{ant}}=&[\mathbf{y}_{1}^{\mathrm{ant}},\cdots ,\mathbf{y}_{V_0}^{\mathrm{ant}}]\nonumber
	\\
	=&\mathbf{H}_{\mathrm{br}}\sum_{k=1}^K{\sqrt{P_k}\mathrm{diag}\{\mathbf{h}_{k,1}\}}\mathbf{E}_0+\mathbf{N}_0\nonumber
	\\
	=&\mathbf{A}_{N_{\mathrm{bs}}}\mathbf{\Lambda A}_{M}^{\mathrm{H}}\sum_{k=1}^K{\sqrt{P_k}\mathrm{diag}\{\mathbf{h}_{k,1}\}}\mathbf{E}_0+\mathbf{N}_0,
\end{align}
where $\mathbf{E}_0=[ \mathbf{e}_{1},\cdots ,\mathbf{e}_{V_0} ] $. Since the number of antennas $N_{\mathrm{bs}}$ is usually large enough, i.e., $N_{\mathrm{bs}}\gg L$, the common AoA matrix $\mathbf{A}_{N_{\mathrm{bs}}}$ in \eqref{Y_0_ant_MIMO} can be efficiently estimated using the DFT-based approach of Algorithm 1 in \cite{9732214}.

\subsection{Stage II: Estimation of common and cascaded channel of user 1}
\begin{figure}[t]
	\centering
	\includegraphics[width=0.48\textwidth]{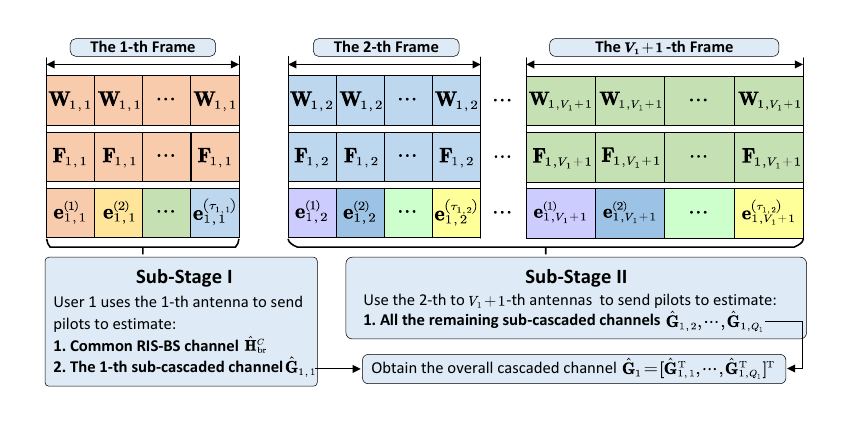}
	\caption{The working diagram of Stage II in the proposed channel estimation protocol.}
	\label{fig_StageII}
\end{figure}

As depicted in Fig. \ref{fig_StageII}, Stage II is composed of two sub-stages in which only user 1 is assumed to send pilot signals. A total of $1+V_1$ ($1+V_1\le Q_1$) frames are allocated to the two sub-stages. In Sub-Stage I, the $1$-st frame is used to estimate the common channel $\mathbf{H}_{\mathrm{br}}^{C}$ and the cascaded channel $\mathbf{G}_{1,1}$ corresponding to the first antenna of user 1. In Sub-Stage II, the remaining $V_1$ frames are used to estimate the cascaded channel for user 1's remaining $Q_1-1$ antennas.

\subsubsection{Sub-Stage I: Estimation of the common and cascaded channel for the $1$-st antenna of a typical user}
In this sub-stage, the $1$-st frame has $\tau _{1,1}$ time slots. User 1 only uses its $1$-st antenna to send pilot signals, which means the hybrid precoding matrix for user 1 is 
\begin{align}
	\mathbf{F}_{1,1}^{\left( j \right)}=\mathbf{F}_{1,1}=\frac{\sqrt{P_1}}{Q_{\mathrm{rf},1}}[ \mathbf{1}_{Q_{\mathrm{rf},1}\times 1},\mathbf{0}_{Q_{\mathrm{rf},1}\times \left( Q_1-1 \right)} ] ^{\mathrm{T}},
\end{align} 
where $1 \le j \le \tau _{1,1}$. For simplicity, we assume that the pilots are chosen as $\mathbf{s}_{1,1}^{\left( 1 \right)}=\cdots =\mathbf{s}_{1,1}^{\left(\tau _{1,1} \right)}=\mathbf{1}_{Q_{\mathrm{rf},1\times 1}}$. Instead of recovering the uncompressed received signals as in Stage I, the hybrid combining matrices are assumed to remain unchanged, i.e. $\mathbf{W}_{1,1}^{\left( 1 \right)}=\cdots =\mathbf{W}_{1,1}^{\left( \tau _{1,1} \right)}=\mathbf{W}_{1,1}$. 

Similar to \eqref{stage1_y}, the received signal at the BS in the $j$-th time slot can be represented as
\begin{align}\label{y_1_1_j_MIMO}
	\mathbf{y}_{1,1}^{\left( j \right)}=&\mathbf{W}_{1,1}^{\left( j \right)}\mathbf{H}_{\mathrm{br}}\mathrm{diag}\{ \mathbf{e}_{1,1}^{\left( j \right)} \} \mathbf{H}_1\mathbf{F}_{1,1}^{\left( j \right)}\mathbf{s}_{1,1}^{\left( j \right)}+\mathbf{W}_{1,1}^{\left( j \right)}\mathbf{n}_{1,1}^{\left( j \right)}\nonumber
	\\
	=&\sqrt{P_1}\mathbf{W}_{1,1}\mathbf{H}_{\mathrm{br}}\mathrm{diag}\left\{ \mathbf{h}_{1,1} \right\} \mathbf{e}_{1,1}^{\left( j \right)}+\mathbf{W}_{1,1}\mathbf{n}_{1,1}^{\left( j \right)}\nonumber
	\\
	=&\sqrt{P_1}\mathbf{W}_{1,1}\mathbf{G}_{1,1}\mathbf{e}_{1,1}^{\left( j \right)}+\mathbf{W}_{1,1}\mathbf{n}_{1,1}^{\left( j \right)},
\end{align}
where $\mathbf{G}_{1,1}=\mathbf{H}_{\mathrm{br}}\mathrm{diag}\left\{ \mathbf{h}_{1,1} \right\} $ is the $1$-st cascaded subchannel of user $1$. Stacking together the received signals in all the $\tau _{1,1}$ time slots, the observations in the $1$-st frame are
\begin{align}\label{Y_1_1_MIMO}
	\mathbf{Y}_{1,1}&=[ \mathbf{y}_{1,1}^{\left( 1 \right)},\cdots ,\mathbf{y}_{1,1}^{\left( \tau _{1,1} \right)} ] \nonumber
	\\
	&=\sqrt{P_1}\mathbf{W}_{1,1}\mathbf{H}_{\mathrm{br}}\mathrm{diag}\{ \mathbf{h}_{1,1} \} \mathbf{E}_{1,1}+\mathbf{W}_{1,1}\mathbf{N}_{1,1},
\end{align}
where $\mathbf{E}_{1,1}=[ \mathbf{e}_{1,1}^{\left( 1 \right)},\cdots ,\mathbf{e}_{1,1}^{\left( \tau _{1,1} \right)} ]$ and $\mathbf{N}_{1,1}=[ \mathbf{n}_{1,1}^{\left( 1 \right)},\cdots ,\mathbf{n}_{1,1}^{\left( \tau _{1,1} \right)} ]$. 

We assume that the number of RF chains always satisfies $N_{\mathrm{rf}}\ge L$, where $L$ is the number of propagation paths from the RIS to the BS. Based on the estimated AoA matrix $\hat{\mathbf{A}}_{\mathrm{bs}}$, the observations $\mathbf{Y}_{1,1}$ are processed as
\begin{align}\label{p_11_p_L1}
	\left[ \mathbf{p}_{1,1},\cdots ,\mathbf{p}_{L,1} \right] =&\frac{1}{\sqrt{P_1}}[ ( \mathbf{W}_{1,1}\hat{\mathbf{A}}_{\mathrm{bs}} ) ^{\dagger}\mathbf{Y}_{1,1} ] ^{\mathrm{H}}\nonumber
	\\
	=&\mathbf{E}_{1,1}^{\mathrm{H}}\mathrm{diag}\{\mathbf{h}_{1,1}^{*}\}\mathbf{A}_M\mathbf{\Lambda }^{\mathrm{H}}+\tilde{\mathbf{N}}_{1,1},
\end{align}
where 
\begin{align}\label{N_w1_1_1}
	\tilde{\mathbf{N}}_{1,1}=\frac{1}{\sqrt{P_1}}\left[ (\mathbf{W}_{1,1}\hat{\mathbf{A}}_{N_{\mathrm{bs}}})^{\dagger}\mathbf{W}_{1,1}\mathbf{N}_{1,1} \right] ^{\mathrm{H}}.
\end{align}
In particular, $\mathbf{p}_{l,1}$ in \eqref{p_11_p_L1} can be written as
\begin{align}\label{p_l_1_ini}
	\mathbf{p}_{l,1}=\mathbf{E}_{1,1}^{\mathrm{H}}\mathrm{diag}\{\mathbf{h}_{1,1}^{*}\}\mathbf{a}_M(\upsilon _l,\omega _l)\alpha _{l}^{*}+\tilde{\mathbf{n}}_{l,1}\nonumber
	\\
	=\mathbf{E}_{1,1}^{\mathrm{H}}\mathrm{diag}\{\mathbf{a}_M(\upsilon _l,\omega _l)\}\mathbf{h}_{1,1}^{*}\alpha _{l}^{*}+\tilde{\mathbf{n}}_{l,1},
\end{align}
where $\tilde{\mathbf{n}}_{l,1}=[ \tilde{\mathbf{N}}_{1,1} ] _{:,l}$. Note that $\mathbf{h}_{1,1}=[ \mathbf{H}_1 ] _{:,1}=\mathbf{A}_{M,1}\boldsymbol{\beta }_1$, so \eqref{p_l_1_ini} can be further represented as
\begin{align}\label{p_l_1}
	\mathbf{p}_{l,1}=&\mathbf{E}_{1,1}^{\mathrm{H}}\mathrm{diag}\{\mathbf{a}_M(\upsilon _l,\omega _l)\}\mathbf{A}_{M,1}^{*}\boldsymbol{\beta }_{1}^{*}\alpha _{l}^{*}+\tilde{\mathbf{n}}_{l,1}\nonumber
	\\
	=&\mathbf{E}_{1,1}^{\mathrm{H}}[\mathbf{a}_M(\upsilon _l-\theta _{1,1},\omega _l-\varphi _{1,1}),\cdots ,\nonumber
	\\
	&.\mathbf{a}_M(\upsilon _l-\theta _{1,J_1},\omega _l-\varphi _{1,J_1})] \boldsymbol{\beta }_{1}^{*}\alpha _{l}^{*}+\tilde{\mathbf{n}}_{l,1}\nonumber
	\\
	=&\mathbf{E}_{1,1}^{\mathrm{H}}\mathbf{A}_{\mathrm{RIS},l}\boldsymbol{\beta }_{\mathrm{RIS},l,1}^{*}+\tilde{\mathbf{n}}_{l,1}\nonumber
	\\
	=&\mathbf{E}_{1,1}^{\mathrm{H}}\mathbf{h}_{\mathrm{RIS},l,1}+\tilde{\mathbf{n}}_{l,1},
\end{align}
where the cascaded gain\footnote{Here, ``1" in $\boldsymbol{\beta }_{\mathrm{RIS},l,1}^{*}$ refers to using the $1$-st antenna to send pilot signals.} $\boldsymbol{\beta }_{\mathrm{RIS},l,1}^{*}$ and the cascaded AoD cosine matrix $\mathbf{A}_{\mathrm{RIS},l}$ are given by
\begin{subequations}
	\begin{align}
		\boldsymbol{\beta }_{\mathrm{RIS},l,1}^{*}=&\boldsymbol{\beta }_{1}^{*}\alpha _{l}^{*},
		\\
		\mathbf{A}_{\mathrm{RIS},l}=&[\mathbf{a}_M(\upsilon _l-\theta _{1,1},\omega _l-\varphi _{1,1}),\nonumber
		\\
		&.\cdots ,\mathbf{a}_M(\upsilon _l-\theta _{1,J_1},\omega _l-\varphi _{1,J_1})],
		\\
		\mathbf{h}_{\mathrm{RIS},l,1}=&\mathbf{A}_{\mathrm{RIS},l}\boldsymbol{\beta }_{\mathrm{RIS},l,1}^{*}.
	\end{align}
\end{subequations}

Denote the index of the typical path as $r$, which is chosen according to the path power:
\begin{align}\label{choose_r}
	r=\mathrm{arg}\underset{1\le i\le L}{\max}\| \mathbf{p}_{i,1} \| _{2}^{2}.
\end{align}
Due to the special structure in \eqref{p_l_1}, $\mathbf{p}_{r,1}$ can be formulated using a $J_1$-sparse signal model, where
\begin{align}\label{cs_p_r_1}
	\mathbf{p}_{r,1}=\mathbf{E}_{1,1}^{\mathrm{H}}( \bar{\mathbf{A}}_1\otimes \bar{\mathbf{A}}_2 ) \mathbf{b}_{r,1}+\tilde{\mathbf{n}}_{r,1},
\end{align}
$\mathbf{b}_{r,1}\in \mathbb{C} ^{D\times 1}$ is a sparse vector with $J_1$ non-zero elements, $\bar{\mathbf{A}}_1\in \mathbb{C} ^{M_1\times D_1}$ and $\bar{\mathbf{A}}_2\in \mathbb{C} ^{M_2\times D_2}$ are over-complete dictionaries with $D_1\ge M_1$ and $D_2\ge M_2$).  Since each column of $\bar{\mathbf{A}}_1$ and $\bar{\mathbf{A}}_2$ represents the array steering vector for possible values of $\upsilon _r-\theta _{1,j_1}\in [ -2\small{\frac{d_{\mathrm{RIS}}}{\lambda}},2\small{\frac{d_{\mathrm{RIS}}}{\lambda}} ]$ and $\omega _r-\varphi _{1,j_1}\in [ -2\small{\frac{d_{\mathrm{RIS}}}{\lambda}},2\small{\frac{d_{\mathrm{RIS}}}{\lambda}} ] \,\,$  respectively, $\bar{\mathbf{A}}_1$ and $\bar{\mathbf{A}}_2$ can be constructed as 
\begin{subequations}\label{A_dic}
	\begin{align}
		\bar{\mathbf{A}}_1=&[\mathbf{a}_{M_1}(-2\small{\frac{d_{\mathrm{RIS}}}{\lambda}}),\cdots ,\mathbf{a}_{M_1}((2-\frac{4}{D_1})\small{\frac{d_{\mathrm{RIS}}}{\lambda}})],
		\\
		\bar{\mathbf{A}}_2=&[\mathbf{a}_{M_2}(-2\small{\frac{d_{\mathrm{RIS}}}{\lambda}}),\cdots ,\mathbf{a}_{M_2}((2-\frac{4}{D_2})\small{\frac{d_{\mathrm{RIS}}}{\lambda}})].
	\end{align}
\end{subequations}

Let $\boldsymbol{\beta }_1=\left[ \beta _{1,1},\cdots ,\beta _{1,J_1} \right] ^{\mathrm{T}}$ and $D=D_1\times D_2$. The $J_1$ non-zero elements of $\mathbf{b}_{r,1}\in \mathbb{C} ^{D\times 1}$ correspond to the $J_1$ cascaded channel gains $\{ \alpha _{l}^{*}\beta _{1,1}^{*},\cdots ,\alpha _{l}^{*}\beta _{1,J_1}^{*} \}$. The parameters of the sparse signal model in \eqref{cs_p_r_1} can be readily found using standard CS-based algorithms such as OMP. However, although OMP will lead to high complexity if it is used in the estimation for other paths $\mathbf{h}_{\mathrm{RIS},l,1}$ for $1\leq l \leq L$, $l\ne r$. Fortunately, the following structure in the model can be exploited to reduce the complexity:
\begin{subequations}
	\begin{align}
		\mathbf{A}_{\mathrm{RIS},l}=&\mathrm{diag}\{\mathbf{a}_M(\upsilon _l-\upsilon _r,\omega _l-\omega _r)\}\mathbf{A}_{\mathrm{RIS},r}\nonumber
		\\
		=&\mathrm{diag}\{\mathbf{a}_M(\Delta \upsilon _l,\Delta \omega _l)\}\mathbf{A}_{\mathrm{RIS},r}\label{A_ris_l_and_A_ris_r}
		\\
		\boldsymbol{\beta }_{\mathrm{RIS},l,1}^{*}=&\boldsymbol{\beta }_{\mathrm{RIS},r,1}^{*}\frac{\alpha _{l}^{*}}{\alpha _{r}^{*}}=\boldsymbol{\beta }_{\mathrm{RIS},r,1}^{*}x_l\label{beta_ris_l_1_and_beta_ris_r_1},
	\end{align}
\end{subequations}
where $x_l=\frac{\alpha _{l}^{*}}{\alpha _{r}^{*}}$ is a gain scaling factor and $\Delta \upsilon _l=\upsilon _l-\upsilon _r$, $\Delta \omega _l=\omega _l-\omega _r$ are angle rotation factors. Thus $\mathbf{h}_{\mathrm{RIS},l,1}$ can be rewritten as
\begin{align}\label{h_RIS_l_1}
	\mathbf{h}_{\mathrm{RIS},l,1}=&\mathbf{A}_{\mathrm{RIS},l}\boldsymbol{\beta }_{\mathrm{RIS},l,1}^{*}\nonumber
	\\
	=&\mathrm{diag}\{\mathbf{a}_M(\Delta \upsilon _l,\Delta \omega _l)\}\mathbf{A}_{\mathrm{RIS},r}\boldsymbol{\beta }_{\mathrm{RIS},r,1}^{*}x_l\nonumber
	\\
	=&\mathrm{diag}\{\mathbf{a}_M(\Delta \upsilon _l,\Delta \omega _l)\}\mathbf{h}_{\mathrm{RIS},r,1}x_l\nonumber
	\\
	=&\mathrm{diag}\{\mathbf{h}_{\mathrm{RIS},r,1}\}\mathbf{a}_M(\Delta \upsilon _l,\Delta \omega _l)x_l.
\end{align}

Once the estimate $\hat{\mathbf{h}}_{\mathrm{RIS},r,1}$ of $\mathbf{h}_{\mathrm{RIS},r,1}$ is found, we can obtain $\hat{\mathbf{h}}_{\mathrm{RIS},l,1}$ by solving a separable nonlinear
least squares (SNL-LS) problem \cite{golub1973differentiation}. After we obtain $(\Delta \hat{\upsilon}_l,\Delta \hat{\omega}_l)$ for $1\leq l\leq L$ and $l\ne r$, $[\hat{\mathbf{h}}_{\mathrm{RIS},1,1},\cdots ,\hat{\mathbf{h}}_{\mathrm{RIS},L,1}]$ can be derived. Thus, the $1$-st cascaded subchannel can be calculated as
\begin{align}\label{hat_G_1_1}
	\hat{\mathbf{G}}_{1,1}=\hat{\mathbf{A}}_{N_{\mathrm{bs}}}[\hat{\mathbf{h}}_{\mathrm{RIS},1,1},\cdots ,\hat{\mathbf{h}}_{\mathrm{RIS},L,1}]^{\mathrm{H}}.
\end{align}
To reduce the pilot overhead for estimating the channels of other users in Stage III, we construct the equivalent common RIS-BS channel \cite{9919846}, which is defined as 
\begin{align}\label{H_br_C}
	\mathbf{H}_{\mathrm{br}}^{C}=\mathbf{A}_{N_{\mathrm{bs}}}\mathbf{\Lambda }^C(\mathbf{A}_{M}^{C})^{\mathrm{H}},
\end{align}
where 
\begin{subequations}
	\begin{align}
		\mathbf{\Lambda }^C=&(\mathbf{1}^{\mathrm{T}}\boldsymbol{\beta }_{1}^{*}\alpha _{r}^{*}\mathrm{diag}\{ x_1,\cdots ,x_L \} ) ^*\nonumber
		\\
		=&\left( \mathbf{1}^{\mathrm{T}}\boldsymbol{\beta }_{\mathrm{RIS},r,1}^{*}\mathrm{diag}\{x_1,\cdots ,x_L\} \right) ^*,\label{large_alpha_C}
		\\
		\mathbf{A}_{M}^{C}=& \mathrm{diag}\{ \mathbf{a}_M( \upsilon _r-\theta ^C,\omega _r-\varphi ^C ) \}\nonumber
		\\
		&\times[\mathbf{a}_M(\Delta \upsilon _1,\Delta \omega _1),\cdots ,\mathbf{a}_M(\Delta \upsilon _L,\Delta \omega _L)],\label{cascaded_AoD_C}
	\end{align}
\end{subequations}
where $\theta ^C=\frac{1}{J_1}\sum_{j_1=1}^{J_1}{\theta _{1,j_1}}$ and $\varphi ^C  =\frac{1}{J_1}\sum_{j_1=1}^{J_1}{\varphi _{1,j_1}}$.
Note that all the parameters in \eqref{large_alpha_C} and \eqref{cascaded_AoD_C} have been obtained during the estimation of $\mathbf{G}_{1,1}$, the equivalent common RIS-BS channel $\hat{\mathbf{H}}_{\mathrm{br}}^{C}=\hat{\mathbf{A}}_{N_{\mathrm{bs}}}\hat{\mathbf{\Lambda}}^C(\hat{\mathbf{A}}_{M}^{C})^{\mathrm{H}}$ can be obtained.

\begin{algorithm}[!t]
	\caption{Estimation of $\mathbf{G}_{1}$, $\hat{\mathbf{\Lambda}}^C$, $\hat{\mathbf{A}}_{M}^{C}$}
	\label{alg_G_1}
	\renewcommand{\algorithmicrequire}{\textbf{Input:}}
	\renewcommand{\algorithmicensure}{\textbf{Output:}}
	\begin{algorithmic}[1]
		\REQUIRE $\mathbf{Y}_{1,i}$, $1 \le i \le V_1+1$.
		\\\textbf{\textit{Sub-Stage I: }}\textit{Estimation of $\mathbf{G}_{1,1}$, $\mathbf{\Lambda}^C$ and $\mathbf{A}_{M}^{C}$}.
		\STATE Calculate $[ \mathbf{p}_{1,1},\cdots ,\mathbf{p}_{L,1} ] =\frac{1}{\sqrt{P_1}}[(\mathbf{W}_{1,1}\hat{\mathbf{A}}_{\mathrm{bs}})^{\dagger}\mathbf{Y}_{1,1}]^{\mathrm{H}}$.
		\STATE Choose the typical path with index $r$ according to \eqref{choose_r}.
		\STATE Find $\hat{\mathbf{h}}_{\mathrm{RIS},r,1}=\hat{\mathbf{A}}_{\mathrm{RIS},r}\hat{\boldsymbol{\beta}}_{\mathrm{RIS},r,1}^{*}$ from \eqref{cs_p_r_1} using the OMP algorithm.
		\FOR{$1 \le l\le L$, $l \ne r$}
		\STATE Find $\Delta \hat{\upsilon}_l$, $\Delta \hat{\omega}_l$ and $\hat{x}_l$ by solving the SNL-LS problem in \eqref{h_RIS_l_1}.
		\STATE Obtain $\hat{\mathbf{A}}_{\mathrm{RIS},l}$ and $\hat{\boldsymbol{\beta}}_{\mathrm{RIS},l,1}^{*}$ based on \eqref{A_ris_l_and_A_ris_r} and \eqref{beta_ris_l_1_and_beta_ris_r_1}.
		\STATE Calculate $\hat{\mathbf{h}}_{\mathrm{RIS},l,1}$ according to \eqref{h_RIS_l_1}.
		\ENDFOR
		\STATE Calculate $\hat{\mathbf{G}}_{1,1}=\hat{\mathbf{A}}_{N_{\mathrm{bs}}}[\hat{\mathbf{h}}_{\mathrm{RIS},1,1},\cdots ,\hat{\mathbf{h}}_{\mathrm{RIS},L,1}]^{\mathrm{H}}$.
		\STATE Calculate $\hat{\mathbf{\Lambda}}^C$ and $\hat{\mathbf{A}}_{M}^{C}$ according to \eqref{large_alpha_C} and \eqref{cascaded_AoD_C}.
		\\\textbf{\textit{Sub-Stage II: }}\textit{Estimation of $\mathbf{G}_{1,q_1}$, $2 \le q_1 \le Q_1$}.
		\FOR{$2 \le i \le V_1+1$}
		\STATE Calculate $[\mathbf{p}_{1,i},\cdots,\mathbf{p}_{L,i}]=\frac{1}{\sqrt{P_1}}[(\mathbf{W}_{1,i}\hat{\mathbf{A}}_{\mathrm{bs}})^{\dagger}\mathbf{Y}_{1,i}]^{\mathrm{H}}$.
		\STATE Calculate $\hat{\boldsymbol{\beta}}_{\mathrm{RIS},r,i}^{*}$ according to \eqref{Pseudo_inverse_p_r_i}.
		\STATE Calculate $\hat{\eta}_{r,j_1,i}^{1}$ according to \eqref{eta_r_j1_i}.
		\ENDFOR
		\STATE Obtain AoDs $\hat{\xi}_{1,j_1}$ for $1 \le j_1 \le J_1$ by solving the one-dimensional search problem in \eqref{one_dimension_searching_for_xi}.
		\STATE Reconstruct the AoD matrix $\hat{\mathbf{A}}_{Q_1}$ based on \eqref{hat_A_Q1}.
		\FOR{$2 \le q_1 \le Q_1$}
		\STATE Calculate $\hat{\mathbf{\Omega}}_{1,q_1}$ according to \eqref{hat_large_omega}.
		\FOR{$1 \le l \le L$}
		\STATE Calculate $\hat{\boldsymbol{\beta}}_{\mathrm{RIS},l,q_1}^{*}$ via \eqref{hat_beta_ris_l_q1}.
		\STATE Calculate $\hat{\mathbf{h}}_{\mathrm{RIS},l,q_1}$ according to \eqref{hat_h_ris_l_q1}.
		\ENDFOR
		\STATE Obtain $\hat{\mathbf{G}}_{1,q_1}$ via \eqref{hat_G_1_q1}.
		\ENDFOR
		\STATE Obtain the overall cascaded channel $\hat{\mathbf{G}}_1$ according to \eqref{overall_hat_G_1}.
		\ENSURE $\hat{\mathbf{G}}_{1}$, $\hat{\mathbf{\Lambda}}^C$, $\hat{\mathbf{A}}_{M}^{C}$.
	\end{algorithmic}
\end{algorithm}

\subsubsection{Sub-Stage II: Estimation of cascaded channel for the typical user with remaining antennas}
In this sub-stage, the remaining $V_1$ frames will be utilized, each of which has $\tau_{1,2}$ time slots. In the $i$-th frame, user 1 transmits pilot signals from its $i$-th antenna. For $2 \le i \le V_1+1$ and $1 \le j \le \tau _{1,2}$, the hybrid precoding matrix at user 1 in the $j$-th time slot of the $i$-th frame can be written as
\begin{align}
	\mathbf{F}_{1,i}^{\left( j \right)}=\mathbf{F}_{1,i}=\frac{\sqrt{P_1}}{Q_{\mathrm{rf},1}}[ \mathbf{0}_{Q_{\mathrm{rf},1}\times (i-1)},\mathbf{1}_{Q_{\mathrm{rf},1}\times 1},\mathbf{0}_{Q_{\mathrm{rf},1}\times ( Q_1-i )} ] ^{\mathrm{T}},
\end{align}
where $1 \le j \le \tau _{1,2}$. We assume that the hybrid combining matrix at the BS remains unchanged, i.e., $\mathbf{W}_{1,i}^{\left( j \right)}=\mathbf{W}_{1,i}$, and the pilot signals are set to $\mathbf{1}_{Q_{\mathrm{rf},1\times 1}}$. Similar to \eqref{Y_1_1_MIMO}, the measurement matrix in the $i$-th frame is
\begin{align}
	\mathbf{Y}_{1,i}=&[\mathbf{y}_{1,i}^{\left( 1 \right)},\cdots ,\mathbf{y}_{1,i}^{\left( \tau _{1,2} \right)}]\nonumber
	\\
	=&\sqrt{P_1}\mathbf{W}_{1,i}\mathbf{H}_{\mathrm{br}}\mathrm{diag}\{ \mathbf{h}_{1,i} \} \mathbf{E}_{1,i}+\mathbf{W}_{1,i}\mathbf{N}_{1,i}.
\end{align}
As in \eqref{p_11_p_L1}, $\mathbf{Y}_{1,i}$ is processed as
\begin{align}
	[\mathbf{p}_{1,i},\cdots,\mathbf{p}_{L,i}]=&\frac{1}{\sqrt{P_1}}[(\mathbf{W}_{1,i}\hat{\mathbf{A}}_{\mathrm{bs}})^{\dagger}\mathbf{Y}_{1,i}]^{\mathrm{H}}\nonumber
	\\
	=&\mathbf{E}_{1,i}^{\mathrm{H}}\mathrm{diag}\{\mathbf{h}_{1,i}^{*}\}\mathbf{A}_M\mathbf{\Lambda }^{\mathrm{H}}+\tilde{\mathbf{N}}_{1,i}.
\end{align}
Note that $\mathbf{h}_{1,i}$ can be written as
\begin{align}\label{h_1_i_and_beta_1}
	\mathbf{h}_{1,i}&=[\mathbf{H}_1]_{:,i}\nonumber
	\\
	&=\mathbf{A}_{M,1}\mathbf{B}_1[ \mathbf{A}_{Q_1}^{\mathrm{H}} ] _{:,i}\nonumber
	\\
	&=\mathbf{A}_{M,1}\mathrm{diag}\{ \boldsymbol{\beta }_1 \} [ \mathbf{A}_{Q_1}^{\mathrm{H}} ] _{:,i}\nonumber
	\\
	&=\mathbf{A}_{M,1}\mathrm{diag}\{ [ \mathbf{A}_{Q_1}^{\mathrm{H}} ] _{:,i} \} \boldsymbol{\beta }_1\nonumber
	\\
	&=\mathbf{A}_{M,1}\mathbf{\Omega }_{1,i}^{*}\boldsymbol{\beta }_1,
\end{align}
where
\begin{align}\label{large_omega}
	\mathbf{\Omega }_{1,i}=&\mathrm{diag}\{ [ \mathbf{A}_{Q_1} ] _{i,:} \}
	\\
	=&\mathrm{diag}\{ e^{-j2\pi (i-1)\xi _{1,1}},\cdots ,e^{-j2\pi (i-1)\xi _{1,J_1}} \}.\nonumber
\end{align}

Similar to the approach taken in Sub-Stage I, $r$ is chosen as the index of a typical path\footnote{Here, the $r$-th path is treated as the typical path with the maximum received power, a choice that improves the accuracy of the subsequent AoD estimates at the user.} and $\mathbf{p}_{r,i}$ can be written as
\begin{align}
	\mathbf{p}_{r,i}=&\mathbf{E}_{1,i}^{\mathrm{H}}\mathrm{diag}\{ \mathbf{h}_{1,i}^{*} \} \mathbf{a}_M( \upsilon _r,\omega _r ) \alpha _{r}^{*}+\tilde{\mathbf{n}}_{r,i}\nonumber
	\\
	=&\mathbf{E}_{1,i}^{\mathrm{H}}\mathrm{diag}\{\mathbf{a}_M(\upsilon _r,\omega _r)\}\mathbf{A}_{M,1}^{*}\mathbf{\Omega }_{1,i}\boldsymbol{\beta }_{1}^{*}\alpha _{r}^{*}+\tilde{\mathbf{n}}_{r,i}\nonumber
	\\
	=&\mathbf{E}_{1,i}^{\mathrm{H}}[ \mathbf{a}_M(\upsilon _r-\theta _{1,1},\omega _r-\varphi _{1,1}),\cdots ,\nonumber
	\\
	&\mathbf{a}_M(\upsilon _r-\theta _{1,J_1},\omega _r-\varphi _{1,J_1}) ]\mathbf{\Omega }_{1,i}\boldsymbol{\beta }_{1}^{*}\alpha _{r}^{*}+\tilde{\mathbf{n}}_{r,i}\nonumber
	\\
	=&\mathbf{E}_{1,i}^{\mathrm{H}}\mathbf{A}_{\mathrm{RIS},r}\boldsymbol{\beta }_{\mathrm{RIS},r,i}^{*}+\tilde{\mathbf{n}}_{r,i}\nonumber
	\\
	=&\mathbf{E}_{1,i}^{\mathrm{H}}\mathbf{h}_{\mathrm{RIS},r,i}+\tilde{\mathbf{n}}_{r,i},
\end{align}
where
\begin{subequations}
	\begin{align}
		\boldsymbol{\beta }_{\mathrm{RIS},r,i}^{*}=&\mathbf{\Omega }_{1,i}\boldsymbol{\beta }_{1}^{*}\alpha _{r}^{*}=\mathbf{\Omega }_{1,i}\boldsymbol{\beta }_{\mathrm{RIS},r,1}^{*}.\label{beta_ris_r_i}
		\\
		\mathbf{h}_{\mathrm{RIS},r,i}=&\mathbf{A}_{\mathrm{RIS},r}\boldsymbol{\beta }_{\mathrm{RIS},r,i}^{*}.\label{h_ris_r_i}
	\end{align}
\end{subequations}
We define
\begin{align}\label{eta_r_j1_i}
	\eta _{r,j_1,i}^{1}=\frac{[ \boldsymbol{\beta }_{\mathrm{RIS},r,i}^{*} ] _{j_1}}{[ \boldsymbol{\beta }_{\mathrm{RIS},r,1}^{*} ] _{j_1}}=e^{-j2\pi ( i-1 ) \xi _{1,j_1}} , 
\end{align}
and stacking these values for $2 \le i \le V_1+1$ yields
\begin{align}
	[ 1,\eta _{r,j_1,2}^{1},\cdots ,\eta _{r,j_1,V_1+1}^{1} ] =\mathbf{a}_{V_1+1}^{\mathrm{T}}( \xi _{1,j_1} ).
\end{align}
Note that $\hat{\mathbf{A}}_{\mathrm{RIS},r}$ has been obtained in Sub-Stage I, and $\boldsymbol{\beta }_{\mathrm{RIS},r,i}^{*}$ is estimated using LS
\begin{align}\label{Pseudo_inverse_p_r_i}
	\hat{\boldsymbol{\beta}}_{\mathrm{RIS},r,i}^{*}=(\mathbf{E}_{1,i}^{\mathrm{H}}\hat{\mathbf{A}}_{\mathrm{RIS},r})^{\dagger}\mathbf{p}_{r,i}=\boldsymbol{\beta }_{\mathrm{RIS},r,i}^{*}+\mathbf{n}_{\beta ,i},
\end{align}
where 
\begin{align}\label{n_w1_i_2}
	\mathbf{n}_{\beta ,i}=(\mathbf{E}_{1,i}^{\mathrm{H}}\hat{\mathbf{A}}_{\mathrm{RIS},r})^{\dagger}\tilde{\mathbf{n}}_{r,i}.
\end{align}
Then $\hat{\eta}_{r,j_1,i}^{1}$ can be obtained based on \eqref{eta_r_j1_i}. 

Using the steering vector in \eqref{ULA_array}, the $j_1$-th AoD of user 1 can be estimated via the following simple one-dimensional search:
\begin{align}\label{one_dimension_searching_for_xi}
	\hat{\xi}_{1,j_1}=\mathrm{arg}\underset{\xi _{1,j_1}}{\max}\,\,\| \mathbf{a}_{V_1+1}^{\mathrm{T}}( \xi _{1,j_1} ) [ 1,\hat{\eta}_{r,j_1,2}^{1},\cdots ,\hat{\eta}_{r,j_1,V_1+1}^{1} ] ^{\mathrm{H}} \| _{2}^{2}.
\end{align}
After the $J_1$ AoDs have been estimated, the AoD matrix of user 1 can be reconstructed as
\begin{align}\label{hat_A_Q1}
	\hat{\mathbf{A}}_{Q_1}=[ \mathbf{a}_{Q_1}(\hat{\xi}_{1,1}),\cdots ,\mathbf{a}_{Q_1}(\hat{\xi}_{1,J_1}) ].
\end{align}
Based on \eqref{large_omega}, we have
\begin{align}\label{hat_large_omega}
	\hat{\mathbf{\Omega}}_{1,q_1}=\mathrm{diag}\{[\hat{\mathbf{A}}_{Q_1}]_{q_1,:}\}, 2\le q_1\le Q_1, 
\end{align}
and then \eqref{beta_ris_r_i}, leads to
\begin{align}\label{hat_beta_ris_l_q1}
	\hat{\boldsymbol{\beta}}_{\mathrm{RIS},l,q_1}^{*}=\hat{\mathbf{\Omega}}_{1,q_1}\hat{\boldsymbol{\beta}}_{\mathrm{RIS},l,1}^{*}, 1 \le l \le L,
\end{align}
where $\hat{\boldsymbol{\beta}}_{\mathrm{RIS},l,1}^{*}$ is derived in Sub-Stage I. According to \eqref{h_ris_r_i}, $\mathbf{h}_{\mathrm{RIS},l,q_1}$ can be estimated as
\begin{align}\label{hat_h_ris_l_q1}
	\hat{\mathbf{h}}_{\mathrm{RIS},l,q_1}=\hat{\mathbf{A}}_{\mathrm{RIS},l}\hat{\boldsymbol{\beta}}_{\mathrm{RIS},l,q_1}^{*},
\end{align}
where $\hat{\mathbf{A}}_{\mathrm{RIS},l}$ is also obtained in Sub-Stage I. As in \eqref{hat_G_1_1}, the $q_1$-th cascaded subchannel is estimated as
\begin{align}\label{hat_G_1_q1}
	\hat{\mathbf{G}}_{1,q_1}=\hat{\mathbf{A}}_{N_{\mathrm{bs}}}[\hat{\mathbf{h}}_{\mathrm{RIS},1,q_1},\cdots ,\hat{\mathbf{h}}_{\mathrm{RIS},L,q_1}]^{\mathrm{H}}.
\end{align}
Finally, the overall cascaded channel of user $1$ can be derived using \eqref{MIMO_G_k_complex}:
\begin{align}\label{overall_hat_G_1}
	\hat{\mathbf{G}}_1=[ \hat{\mathbf{G}}_{1,1}^{\mathrm{T}},\cdots ,\hat{\mathbf{G}}_{1,Q_1}^{\mathrm{T}} ] ^{\mathrm{T}}.
\end{align}

\subsection{Stage III: Cascaded channel estimation for remaining users}
In this stage, the remaining $K-1$ users send pilot signals sequentially to estimate their cascade channels $\mathbf{G}_k$, respectively. The estimation process is similar for all the remaining users and we use user $k$ for example to introduce the estimation scheme in the following. Similar to Stage II, Stage III is divided into two sub-stages and a total of $1+V_k$ frames are used. In the first sub-stage, user $k$ only uses the $1$-st antenna to send pilot signals in the $1$-st frame for the estimation of the $1$-st cascaded subchannel. In the second sub-stage, the remaining $V_k$ frames are utilized to estimate all the remaining $Q_k-1$ cascaded subchannels.

\subsubsection{Sub-Stage I: Estimation of the $1$-st cascaded subchannel $\mathbf{G}_{k,1}$}
In this sub-stage, the $1$-st frame has $\tau _{k,1}$ time slots. In each time slot, user $k$ only uses the $1$-st antenna to send a pilot signal, which means the hybrid precoding matrix at user $k$ is 
\begin{align}
	\mathbf{F}_{k,1}^{\left( j \right)}=\mathbf{F}_{k,1}=\frac{\sqrt{P_k}}{Q_{\mathrm{rf},k}}[ \mathbf{1}_{Q_{\mathrm{rf},k}\times 1},\mathbf{0}_{Q_{\mathrm{rf},k}\times \left( Q_k-1 \right)} ] ^{\mathrm{T}},
\end{align}
where $1 \le j \le \tau _{k,1}$. For simplicity, we also set the pilot signals to $\mathbf{1}_{Q_{\mathrm{rf},k\times 1}}$, i.e., $\mathbf{s}_{k,1}^{\left( 1 \right)}=\cdots =\mathbf{s}_{k,1}^{\left(\tau _{k,1} \right)}=\mathbf{1}_{Q_{\mathrm{rf},1\times 1}}$, and the hybrid combining matrix remains unchanged, i.e., $\mathbf{W}_{k,1}^{\left( 1 \right)}=\cdots =\mathbf{W}_{k,1}^{\left( \tau _{k,1} \right)}=\mathbf{W}_{k,1}$. Similar to \eqref{Y_1_1_MIMO},
the measurement matrix in the $1$-st frame can be written as
\begin{align}\label{Y_k_1_MIMO}
	&\mathbf{Y}_{k,1}=[\mathbf{y}_{k,1}^{\left( 1 \right)},\cdots ,\mathbf{y}_{k,1}^{\left( \tau _{k,1} \right)}]\nonumber
	\\
	&=\sqrt{P_k}\mathbf{W}_{k,1}\mathbf{H}_{\mathrm{br}}\mathrm{diag}\left\{ \mathbf{h}_{k,1} \right\} \mathbf{E}_{k,1}+\mathbf{W}_{k,1}\mathbf{N}_{k,1}
	\\
	&=\sqrt{P_k}\mathbf{W}_{k,1}\mathbf{A}_{N_{\mathrm{bs}}}\mathbf{\Lambda A}_{M}^{\mathrm{H}}\mathrm{diag}\left\{ \mathbf{h}_{k,1} \right\} \mathbf{E}_{k,1}+\mathbf{W}_{k,1}\mathbf{N}_{k,1}.\nonumber
\end{align} 
Based on $\hat{\mathbf{A}}_{\mathrm{bs}}$ obtained in Stage II, the measurement matrix in \eqref{Y_k_1_MIMO} can be processed similar to \eqref{p_11_p_L1}:
\begin{align}\label{pse_Y_k_1}
	\tilde{\mathbf{Y}}_{k,1}=&\frac{1}{\sqrt{P_k}}(\mathbf{W}_{k,1}\hat{\mathbf{A}}_{\mathrm{bs}})^{\dagger}\mathbf{Y}_{k,1}\nonumber
	\\
	=&\mathbf{\Lambda A}_{M}^{\mathrm{H}}\mathrm{diag}\{ \mathbf{h}_{k,1} \} \mathbf{E}_{k,1}+\tilde{\mathbf{N}}_{k,1},
\end{align}
where the noise term is
\begin{align}\label{tilde_N_k_1}
	&\tilde{\mathbf{N}}_{k,1}=\frac{1}{\sqrt{P_k}}(\mathbf{W}_{k,1}\hat{\mathbf{A}}_{\mathrm{bs}})^{\dagger}\mathbf{W}_{k,1}\mathbf{N}_{k,1}.
\end{align}

Using \eqref{large_alpha_C} and \eqref{cascaded_AoD_C}, \eqref{pse_Y_k_1} can be rewritten as
\begin{align}\label{pse_Y_k_1_C}
	\tilde{\mathbf{Y}}_{k,1}=&\mathbf{\Lambda A}_{M}^{\mathrm{H}}\mathrm{diag}\left\{ \mathbf{h}_{k,1} \right\} \mathbf{E}_{k,1}+\tilde{\mathbf{N}}_{k,1}\nonumber
	\\
	=&\frac{1}{\mathbf{1}^{\mathrm{T}}\boldsymbol{\beta }_1}\mathbf{\Lambda }^C(\mathbf{A}_{M}^{C})^{\mathrm{H}}\mathrm{diag}\{\mathbf{a}_M(-\theta ^C,-\varphi ^C)\}\nonumber
	\\
	&\times\mathrm{diag}\{\mathbf{h}_{k,1}\}\mathbf{E}_{k,1}+\tilde{\mathbf{N}}_{k,1}\nonumber
	\\
	=&\mathbf{\Lambda }^C(\mathbf{A}_{M}^{C})^{\mathrm{H}}\mathrm{diag}\{\mathrm{diag}\{\mathbf{a}_M(-\theta ^C,-\varphi ^C)\}\mathbf{h}_{k,1}\}\nonumber
	\\
	&\times\frac{1}{\mathbf{1}^{\mathrm{T}}\boldsymbol{\beta }_1}\mathbf{E}_{k,1}+\tilde{\mathbf{N}}_{k,1}\nonumber
	\\
	=&\mathbf{\Lambda }^C(\mathbf{A}_{M}^{C})^{\mathrm{H}}\mathrm{diag}\{\mathbf{h}_{k,1}^{C}\}\mathbf{E}_{k,1}+\tilde{\mathbf{N}}_{k,1},
\end{align}
where $\mathbf{h}_{k,1}^{C}$ is the channel from user $k$ to the RIS when using the $1$-st antenna, which can be represented as
\begin{align}\label{h_k_1_C_sparse}
	\mathbf{h}_{k,1}^{C}=&\mathrm{diag}\{\mathbf{a}_M(-\theta ^C,-\varphi ^C)\}\mathbf{h}_{k,1}\frac{1}{\mathbf{1}^{\mathrm{T}}\boldsymbol{\beta }_1}\nonumber
	\\
	=&\mathrm{diag}\{\mathbf{a}_M(-\theta ^C,-\varphi ^C)\}\mathbf{A}_{M,k}\boldsymbol{\beta }_k\frac{1}{\mathbf{1}^{\mathrm{T}}\boldsymbol{\beta }_1}\nonumber
	\\
	=&\mathbf{A}_{M,k}^{C}\boldsymbol{\beta }_{k,1}^{C},
\end{align}
where 
\begin{subequations}
	\begin{align}
		\mathbf{A}_{M,k}^{C}=&\mathrm{diag}\{\mathbf{a}_M(-\theta ^C,-\varphi ^C)\}\mathbf{A}_{M,k}\label{A_M_k_C},
		\\
		\boldsymbol{\beta }_{k,1}^{C}=&\boldsymbol{\beta }_k\frac{1}{\mathbf{1}^{\mathrm{T}}\boldsymbol{\beta }_1}.
	\end{align}
\end{subequations}

Exploiting the property $\mathrm{vec}( \mathbf{X}\mathrm{diag}\{ \mathbf{h} \} \mathbf{Y} ) =( \mathbf{Y}^{\mathrm{T}}\diamond \mathbf{X} ) \mathbf{h}$, the vectorization of $\tilde{\mathbf{Y}}_{k,1}$ in \eqref{pse_Y_k_1_C} can be written as
\begin{align}\label{vec_Y_k_1_MIMO}
	\mathrm{vec(}\tilde{\mathbf{Y}}_{k,1})=&(\mathbf{E}_{k,1}^{\mathrm{T}}\diamond (\mathbf{\Lambda }^C(\mathbf{A}_{M}^{C})^{\mathrm{H}}))\mathbf{h}_{k,1}^{C}+\mathrm{vec(}\tilde{\mathbf{N}}_{k,1}).
\end{align}
Replacing $\mathbf{h}_{k,1}^{C}$ with $\mathbf{h}_{k,1}^{C}=\mathbf{A}_{M,k}^{C}\boldsymbol{\beta }_{k,1}^{C}$ from \eqref{h_k_1_C_sparse}, $\mathrm{vec(}\tilde{\mathbf{Y}}_{k,1})$ in \eqref{vec_Y_k_1_MIMO} can be approximated using the virtual angular domain (VAD) representation 
\begin{align}\label{vec_Y_k_1_MIMO_VAD}
	&\mathrm{vec(}\tilde{\mathbf{Y}}_{k,1})=(\mathbf{E}_{k,1}^{\mathrm{T}}\diamond (\mathbf{\Lambda }^C(\mathbf{A}_{M}^{C})^{\mathrm{H}}))\mathbf{A}_{M,k}^{C}\boldsymbol{\beta }_{k,1}^{C}+\mathrm{vec(}\tilde{\mathbf{N}}_{k,1})\nonumber
	\\
	&=(\mathbf{E}_{k,1}^{\mathrm{T}}\diamond (\mathbf{\Lambda }^C(\mathbf{A}_{M}^{C})^{\mathrm{H}}))( \bar{\mathbf{A}}_1\otimes \bar{\mathbf{A}}_2 ) \mathbf{b}_{k,1}+\mathrm{vec(}\tilde{\mathbf{N}}_{k,1}),
\end{align} 
where $\bar{\mathbf{A}}_1$ and $\bar{\mathbf{A}}_2$ have been defined in \eqref{A_dic} and $\mathbf{b}_{k,1}\in \mathbb{C} ^{D\times 1}$ is a sparse vector with $J_k$ nonzero elements $\{ \frac{1}{\mathbf{1}^{\mathrm{T}}\boldsymbol{\beta }_1}\beta _{k,1},\cdots ,\frac{1}{\mathbf{1}^{\mathrm{T}}\boldsymbol{\beta }_1}\beta _{k,J_k} \}$. Define the dictionary $\mathbf{R}_k$ as
\begin{align}\label{R_k}
	\mathbf{R}_k=(\mathbf{E}_{k,1}^{\mathrm{T}}\diamond (\mathbf{\Lambda }^C(\mathbf{A}_{M}^{C})^{\mathrm{H}}))( \bar{\mathbf{A}}_1\otimes \bar{\mathbf{A}}_2 ),
\end{align}
so that $\mathrm{vec(}\tilde{\mathbf{Y}}_{k,1})$ in \eqref{vec_Y_k_1_MIMO_VAD} can be written in the form of a sparse signal recovery problem
\begin{align}\label{vec_Y_k_1_MIMO_VAD_concise}
	\mathrm{vec(}\tilde{\mathbf{Y}}_{k,1})=\mathbf{R}_k\mathbf{b}_{k,1}+\mathrm{vec(}\tilde{\mathbf{N}}_{k,1}),
\end{align}
which can be readily solved by CS techniques. 

According to \eqref{pse_Y_k_1_C}, we have
\begin{align}
	\mathbf{\Lambda A}_{M}^{\mathrm{H}}\mathrm{diag}\{\mathbf{h}_{k,1}\}=\mathbf{\Lambda }^C(\mathbf{A}_{M}^{C})^{\mathrm{H}}\mathrm{diag}\{\mathbf{h}_{k,1}^{C}\}.
\end{align}
Thus, the $1$-st cascaded subchannel $\mathbf{G}_{k,1}$ can be rewritten as
\begin{align}\label{G_k_1_equivalent}
	\mathbf{G}_{k,1}=&\mathbf{H}_{\mathrm{br}}\mathrm{diag}\{ \mathbf{h}_{k,1} \}\nonumber
	\\
	=&\mathbf{A}_{N_{\mathrm{bs}}}\mathbf{\Lambda A}_{M}^{\mathrm{H}}\mathrm{diag}\{ \mathbf{h}_{k,1} \}\nonumber
	\\
	=&\mathbf{A}_{N_{\mathrm{bs}}}\mathbf{\Lambda }^C(\mathbf{A}_{M}^{C})^{\mathrm{H}}\mathrm{diag}\{\mathbf{h}_{k,1}^{C}\},
\end{align}
where $\mathbf{\Lambda }^C$ and $\mathbf{A}_{M}^{C}$ are estimated in Stage II. Denoting $\hat{\mathbf{h}}_{k,1}^{C}$ as the estimate of $\mathbf{h}_{k,1}^{C}$, $\mathbf{G}_{k,1}$ can be found using \eqref{G_k_1_equivalent}.

\begin{algorithm}[t]
	\caption{Estimation of $\mathbf{G}_{k}$}
	\label{alg_G_k}
	\renewcommand{\algorithmicrequire}{\textbf{Input:}}
	\renewcommand{\algorithmicensure}{\textbf{Output:}}
	\begin{algorithmic}[1]
		\REQUIRE $\mathbf{Y}_{k,i}$, $1 \le i \le V_k+1$.
		\STATE Return $\hat{\mathbf{\Lambda}}^C$ and $\hat{\mathbf{A}}_{M}^{C}$ from Algorithm \ref{alg_G_1}.
		\\\textbf{\textit{Sub-Stage I: }}\textit{Estimation of $\mathbf{G}_{k,1}$}.
		\STATE Calculate $\tilde{\mathbf{Y}}_{k,1}=\frac{1}{\sqrt{P_k}}(\mathbf{W}_{k,1}\hat{\mathbf{A}}_{\mathrm{bs}})^{\dagger}\mathbf{Y}_{k,1}$.
		\STATE Find $\hat{\mathbf{h}}_{k,1}^{C}=\hat{\mathbf{A}}_{M,k}^{C}\hat{\boldsymbol{\beta}}_{k,1}^{C}$ from \eqref{vec_Y_k_1_MIMO_VAD_concise} using the OMP algorithm.
		\\\textbf{\textit{Sub-Stage II: }}\textit{Estimation of $\mathbf{G}_{k,q_k}$, $2 \le q_k \le Q_k$}.
		\FOR{$2 \le i \le V_k+1$}
		\STATE Calculate $\tilde{\mathbf{Y}}_{k,i}=\frac{1}{\sqrt{P_k}}(\mathbf{W}_{k,i}\hat{\mathbf{A}}_{\mathrm{bs}})^{\dagger}\mathbf{Y}_{k,i}$.
		\STATE Calculate the pseudo inverse matrix \newline $	\mathbf{\Pi }^{\dagger}=( ( \mathbf{E}_{k,i}^{\mathrm{T}}\diamond ( \hat{\mathbf{\Lambda}}^C( \hat{\mathbf{A}}_{M}^{C} ) ^{\mathrm{H}} ) ) \hat{\mathbf{A}}_{M,k}^{C} ) ^{\dagger}$.
		\STATE Calculate $\hat{\boldsymbol{\beta}}_{k,i}^{C}$ via \eqref{hat_beta_k_i_C}.
		\STATE Calculate $\hat{\eta}_{j_k,i}^{k}$ according to \eqref{eta_k_jk_i}.
		\ENDFOR
		\STATE Obtain AoDs $\hat{\xi}_{k,j_k}$ for $1 \le j_k \le J_k$ by solving the one-dimensional search problem in \eqref{one_dimension_searching_for_xi_k}.
		\STATE Reconstruct the AoD matrix $\hat{\mathbf{A}}_{Q_k}$ according to \eqref{hat_A_Qk}.
		\FOR{$2 \le q_k \le Q_k$}
		\STATE Calculate $\hat{\mathbf{\Omega}}_{k,q_k}$ according to \eqref{hat_large_omega_k_i}.
		\STATE Calculate $\hat{\boldsymbol{\beta}}_{k,q_k}^{C}$ via \eqref{hat_beta_k_qk_C}.
		\STATE Calculate $\hat{\mathbf{h}}_{k,q_k}^{C}$ according to \eqref{hat_h_k_qk_C}.
		\STATE Obtain $\hat{\mathbf{G}}_{k,q_k}$ via \eqref{hat_G_k_qk}.
		\ENDFOR
		\ENSURE $\hat{\mathbf{G}}_k=[ \hat{\mathbf{G}}_{k,1}^{\mathrm{T}},\cdots ,\hat{\mathbf{G}}_{k,Q_k}^{\mathrm{T}} ] ^{\mathrm{T}}$.
	\end{algorithmic}
\end{algorithm}

\subsubsection{Sub-Stage II: Estimation of the remaining cascaded subchannels of user $k$}
In this sub-stage, similar to Sub-Stage II of Stage II, the remaining $V_k$ frames will be used, each frame containing $\tau_{k,2}$ time slots. In the $i$-th
frame, user $k$ uses the $i$-th antenna to send pilot signals, and the hybrid precoding matrix at user $k$ in the $j$-th time slot of the $i$-th frame is
\begin{align}
	\mathbf{F}_{k,i}^{\left( j \right)}=\mathbf{F}_{k,i}=\frac{\sqrt{P_k}}{Q_{\mathrm{rf},k}}[\mathbf{0}_{Q_{\mathrm{rf},k}\times (i-1)},\mathbf{1}_{Q_{\mathrm{rf},k}\times 1},\mathbf{0}_{Q_{\mathrm{rf},k}\times (Q_k-i)}]^{\mathrm{T}},
\end{align}
where $1 \le j \le \tau _{k,2}$. The hybrid combining matrix at the BS still remains unchanged, i.e., $\mathbf{W}_{i}^{\left( j \right)}=\mathbf{W}_{k,i}$, and the pilot signals are set to $\mathbf{1}_{Q_{\mathrm{rf},k\times 1}}$ for simplicity. 

The measurement matrix in the $i$-th frame is processed as in \eqref{pse_Y_k_1},
\begin{align}\label{pse_Y_k_i}
	\tilde{\mathbf{Y}}_{k,i}=&\frac{1}{\sqrt{P_k}}(\mathbf{W}_{k,i}\hat{\mathbf{A}}_{\mathrm{bs}})^{\dagger}\mathbf{Y}_{k,i}\nonumber
	\\
	=&\mathbf{\Lambda A}_{M}^{\mathrm{H}}\mathrm{diag}\{ \mathbf{h}_{k,i} \} \mathbf{E}_{k,i}+\tilde{\mathbf{N}}_{k,i},
\end{align}
and similar to \eqref{pse_Y_k_1_C}, \eqref{large_alpha_C} and \eqref{cascaded_AoD_C} are used in \eqref{pse_Y_k_i} to obtain
\begin{align}\label{pse_Y_k_i_C}
	\tilde{\mathbf{Y}}_{k,i}=&\mathbf{\Lambda A}_{M}^{\mathrm{H}}\mathrm{diag}\{\mathbf{h}_{k,i}\}\mathbf{E}_{k,i}+\tilde{\mathbf{N}}_{k,i}\nonumber
	\\
	=&\mathbf{\Lambda }^C(\mathbf{A}_{M}^{C})^{\mathrm{H}}\mathrm{diag}\{\mathbf{h}_{k,i}^{C}\}\mathbf{E}_{k,i}+\tilde{\mathbf{N}}_{k,i},
\end{align}
where $\mathbf{h}_{k,i}^{C}$ is the channel from user $k$ to the RIS corresponding to the $i$-th antenna. 

Like \eqref{h_1_i_and_beta_1}, $\mathbf{h}_{k,i}$ can be represented as
\begin{align}\label{h_k_i_and_beta_k}
	\mathbf{h}_{k,i}=[\mathbf{H}_k]_{:,i}=\mathbf{A}_{M,k}\mathbf{\Omega }_{k,i}^{*}\boldsymbol{\beta }_k,
\end{align}
where 
\begin{align}\label{large_omega_k_i}
	\mathbf{\Omega }_{k,i}&=\mathrm{diag}\{ [ \mathbf{A}_{Q_k} ] _{i,:} \}
	\\
	&=\mathrm{diag}\{ e^{-j2\pi (i-1)\xi _{k,1}},\cdots ,e^{-j2\pi (i-1)\xi _{k,J_k}} \}.\nonumber
\end{align}
Thus, following \eqref{h_k_1_C_sparse}, we have 
\begin{align}\label{h_k_i_C_sparse}
	\mathbf{h}_{k,i}^{C}=&\mathrm{diag}\{\mathbf{a}_M(-\theta ^C,-\varphi ^C)\}\mathbf{h}_{k,i}\frac{1}{\mathbf{1}^{\mathrm{T}}\boldsymbol{\beta }_1}\nonumber
	\\
	=&\mathrm{diag}\{\mathbf{a}_M(-\theta ^C,-\varphi ^C)\}\mathbf{A}_{M,k}\mathbf{\Omega }_{k,i}^{*}\boldsymbol{\beta }_k\frac{1}{\mathbf{1}^{\mathrm{T}}\boldsymbol{\beta }_1}\nonumber
	\\
	=&\mathbf{A}_{M,k}^{C}\boldsymbol{\beta }_{k,i}^{C},
\end{align}
where $\mathbf{A}_{M,k}^{C}$ is defined in \eqref{A_M_k_C} and $\boldsymbol{\beta }_{k,i}^{C}$ is
\begin{align}\label{beta_k_i_C}
	\boldsymbol{\beta }_{k,i}^{C}=\mathbf{\Omega }_{k,i}^{*}\boldsymbol{\beta }_k\frac{1}{\mathbf{1}^{\mathrm{T}}\boldsymbol{\beta }_1}=\mathbf{\Omega }_{k,i}^{*}\boldsymbol{\beta }_{k,1}^{C}.
\end{align}
As in \eqref{vec_Y_k_1_MIMO}, we convert \eqref{pse_Y_k_i_C} to vector form,
\begin{align}\label{vec_Y_k_i_MIMO}
	\mathrm{vec(}\tilde{\mathbf{Y}}_{k,i})&=(\mathbf{E}_{k,i}^{\mathrm{T}}\diamond (\mathbf{\Lambda }^C(\mathbf{A}_{M}^{C})^{\mathrm{H}}))\mathbf{h}_{k,i}^{C}+\mathrm{vec(}\tilde{\mathbf{N}}_{k,i})
	\\
	&=(\mathbf{E}_{k,i}^{\mathrm{T}}\diamond (\mathbf{\Lambda }^C(\mathbf{A}_{M}^{C})^{\mathrm{H}}))\mathbf{A}_{M,k}^{C}\boldsymbol{\beta }_{k,i}^{C}+\mathrm{vec(}\tilde{\mathbf{N}}_{k,i}).\nonumber
\end{align}

Using \eqref{beta_k_i_C}, similar to \eqref{eta_r_j1_i} we define
\begin{align}\label{eta_k_jk_i}
	\eta _{j_k,i}^{k}=\frac{[ \boldsymbol{\beta }_{k,i}^{C} ] _{j_k}}{[ \boldsymbol{\beta }_{k,1}^{C} ] _{j_k}}=e^{j2\pi \left( i-1 \right) \xi _{k,j_k}},
\end{align}
and stack them to form the vector
\begin{align}
	[ 1,\eta _{j_k,2}^{k},\cdots ,\eta _{j_k,V_k+1}^{k} ] =\mathbf{a}_{V_k+1}^{\mathrm{H}}( \xi _{k,j_k} ).
\end{align}
Using
\begin{align}\label{Pi_dagger}
	\mathbf{\Pi }^{\dagger}=( ( \mathbf{E}_{k,i}^{\mathrm{T}}\diamond ( \hat{\mathbf{\Lambda}}^C( \hat{\mathbf{A}}_{M}^{C} ) ^{\mathrm{H}} ) ) \hat{\mathbf{A}}_{M,k}^{C} ) ^{\dagger},
\end{align}
$\boldsymbol{\beta }_{k,i}^{C}$ in \eqref{vec_Y_k_i_MIMO} is estimated by
\begin{align}\label{hat_beta_k_i_C}
	\hat{\boldsymbol{\beta}}_{k,i}^{C}=\mathbf{\Pi }^{\dagger}\mathrm{vec(}\tilde{\mathbf{Y}}_{k,i})=\boldsymbol{\beta }_{k,i}^{C}+\mathbf{\Pi }^{\dagger}\mathrm{vec(}\tilde{\mathbf{N}}_{k,i}),
\end{align}
and $\eta _{j_k,i}^{k}$ is estimated based on \eqref{eta_k_jk_i}. 

Similar to \eqref{one_dimension_searching_for_xi} in Stage II, the $j_k$-th AoD of user $k$ can be estimated via a simple one-dimensional search 
\begin{align}\label{one_dimension_searching_for_xi_k}
	\hat{\xi}_{k,j_k}=\mathrm{arg}\underset{\xi _{k,j_k}}{\max}\,\,\| \mathbf{a}_{V_k+1}^{\mathrm{T}}( \xi _{k,j_k} ) [ 1,\hat{\eta}_{j_k,2}^{k},\cdots ,\hat{\eta}_{j_k,V_k+1}^{k} ] ^{\mathrm{T}} \| _{2}^{2}.
\end{align}
After the $J_k$ AoDs have been estimated, the AoD matrix of user $k$ can be reconstructed as
\begin{align}\label{hat_A_Qk}
	\hat{\mathbf{A}}_{Q_k}=[ \mathbf{a}_{Q_k}(\hat{\xi}_{k,1}),\cdots ,\mathbf{a}_{Q_k}(\hat{\xi}_{k,J_k}) ],
\end{align}
and based on \eqref{large_omega_k_i} we obtain
\begin{align}\label{hat_large_omega_k_i}
	\hat{\mathbf{\Omega}}_{k,q_k}=\mathrm{diag}\{[\hat{\mathbf{A}}_{Q_k}]_{q_k,:}\}, 2\le q_k\le Q_k.
\end{align}
Using \eqref{beta_k_i_C}, $\hat{\boldsymbol{\beta}}_{k,q_k}^{C}$ is calculated as
\begin{align}\label{hat_beta_k_qk_C}
	\hat{\boldsymbol{\beta}}_{k,q_k}^{C}=\hat{\mathbf{\Omega}}_{k,q_k}^{*}\hat{\boldsymbol{\beta}}_{k,1}^{C}
\end{align}
for $2 \le q_k \le Q_k$,
where $\hat{\boldsymbol{\beta}}_{k,1}^{C}$ is derived in Sub-Stage I. Then, $\mathbf{h}_{k,q_k}^{C}$ is estimated via \eqref{h_k_i_C_sparse} as
\begin{align}\label{hat_h_k_qk_C}
	\hat{\mathbf{h}}_{k,q_k}^{C}=\hat{\mathbf{A}}_{M,k}^{C}\hat{\boldsymbol{\beta}}_{k,q_k}^{C},
\end{align}
where $\hat{\mathbf{A}}_{M,k}^{C}$ is also found in Sub-Stage I. As in \eqref{G_k_1_equivalent}, the $k$-th cascaded subchannel is estimated as
\begin{align}\label{hat_G_k_qk}
	\hat{\mathbf{G}}_{k,q_k}=\hat{\mathbf{A}}_{N_{\mathrm{bs}}}\hat{\mathbf{\Lambda}}^C(\hat{\mathbf{A}}_{M}^{C})^{\mathrm{H}}\mathrm{diag}\{\hat{\mathbf{h}}_{k,q_k}^{C}\},
\end{align}
and the overall cascaded channel for user $k$ is given by
\begin{align}
	\hat{\mathbf{G}}_k=[ \hat{\mathbf{G}}_{k,1}^{\mathrm{T}},\cdots ,\hat{\mathbf{G}}_{k,Q_k}^{\mathrm{T}} ] ^{\mathrm{T}}.
\end{align}

\subsection{Optimization of hybrid combiner $\mathbf{W}$ and RIS phase shifts $\mathbf{E}$}
\subsubsection{Optimization for hybrid combining matrix $\mathbf{W}$}
The noise powers in Sub-Stage I of Stages II and III are written as \eqref{N_w1_1_1} and \eqref{tilde_N_k_1}, both of which are functions of the hybrid combining matrix $\mathbf{W}$. We have the following proposition.

\begin{proposition}\label{proposition1}
	Define the singular value decomposition (SVD) of $\mathbf{W}\hat{\mathbf{A}}_{N_\mathrm{bs}}$ as
	\begin{align}\label{SVD_W_1_1}
		\mathbf{W}\hat{\mathbf{A}}_{N_\mathrm{bs}}=\mathbf{U}[ \begin{matrix}
			\mathbf{\Lambda }&		\mathbf{0}_{L\times \left( N_{\mathrm{rf}}-L \right)}\\
		\end{matrix} ] ^{\mathrm{T}}\mathbf{V}^{\mathrm{H}}.
	\end{align}
	Assuming that $(\mathbf{W}\hat{\mathbf{A}}_{N_{\mathrm{bs}}})^{\dagger}$ exists, if the hybrid combining matrix $\mathbf{W}$ satisfies
	\begin{align}
		\mathbf{W}\mathbf{W}^{\mathrm{H}}\succeq \lambda _{\max}\mathbf{I}_L,
	\end{align}
	where $\lambda _{\max}$ is the largest element in $\mathrm{diag}\{\mathbf{\Lambda }^{\mathrm{H}}\mathbf{\Lambda }\} $, the noise powers in \eqref{p_l_1} and \eqref{tilde_N_k_1} respectively satisfy
    \begin{subequations}
        \begin{align}
		[ \mathbf{C}_{\tilde{\mathbf{n}}_{l,1}} ] _{i,i}\geq & P_{1}^{-1}\sigma ^2 \\
		[\mathbf{C}_{[ \tilde{\mathbf{N}}_{k,1} ] _{:,l}} ] _{i,i}\geq & P_{k}^{-1}\sigma ^2.
	\end{align}
    \end{subequations}
	\begin{proof}
		Please refer to Appendix A. 
	\end{proof}

\end{proposition}

Based on Proposition \ref{proposition1}, we find that $\mathbf{W}$ must be carefully designed to reduce the effective noise power. Below we propose a novel construction of the hybrid combing matrix for Stages II and III,
\begin{align}\label{W_opt}
	\mathbf{W}_{A}=\left[ \begin{array}{c}
		\hat{\mathbf{A}}_{N_{\mathrm{bs}}}^{\mathrm{H}}\\
		\mathbf{0}_{( N_{\mathrm{rf}}-L ) \times N_{\mathrm{bs}}}\\
	\end{array} \right] ,
\end{align}
where $\hat{\mathbf{A}}_{N_{\mathrm{bs}}}$ is derived in Stage I. 

In Stage II, let $\mathbf{W}_{1,1}=\mathbf{W}_{A}$ so that
\begin{align}\label{W_1_1_mpl_AoAm}
	\mathbf{W}_{A}\hat{\mathbf{A}}_{N_{\mathrm{bs}}}=&\left[ \begin{array}{c}
		\hat{\mathbf{A}}_{N_{\mathrm{bs}}}^{\mathrm{H}}\hat{\mathbf{A}}_{N_{\mathrm{bs}}}\\
		\mathbf{0}_{\left( N_{\mathrm{rf}}-L \right) \times L}\\
	\end{array} \right] 
	=\left[ \begin{array}{c}
		N_{\mathrm{bs}}\mathbf{I}_L\\
		\mathbf{0}_{\left( N_{\mathrm{rf}}-L \right) \times L}\\
	\end{array} \right] .
\end{align}
The noise term in \eqref{N_w1_1_1} can be rewritten as
\begin{align}\label{N_w1_1_1_opt}
	\tilde{\mathbf{N}}_{1,1}=&\frac{1}{\sqrt{P_1}}[ (\mathbf{W}_{1,1}\hat{\mathbf{A}}_{N_{\mathrm{bs}}})^{\dagger}\mathbf{W}_{1,1}\mathbf{N}_{1,1} ] ^{\mathrm{H}}\nonumber
	\\
	=&\frac{1}{\sqrt{P_1}}[ N_{\mathrm{bs}}^{-2}[ \begin{matrix}
		N_{\mathrm{bs}}\mathbf{I}_L&		\mathbf{0}_{L\times \left( N_{\mathrm{rf}}-L \right)}\\
	\end{matrix} ] \mathbf{W}_{A}\mathbf{N}_{1,1} ] ^{\mathrm{H}}\nonumber
	\\
	=&(\sqrt{P_1}N_{\mathrm{bs}})^{-1}\mathbf{N}_{1,1}^{\mathrm{H}}\mathbf{W}_{A}^{\mathrm{H}}[ \begin{matrix}
		\mathbf{I}_L&		\mathbf{0}_{L\times \left( N_{\mathrm{rf}}-L \right)}\\
	\end{matrix} ] ^{\mathrm{H}}\nonumber
	\\
	=&(\sqrt{P_1}N_{\mathrm{bs}})^{-1}\mathbf{N}_{1,1}^{\mathrm{H}}\hat{\mathbf{A}}_{N_{\mathrm{bs}}}.
\end{align}
Note that $[ \mathbf{N}_{1,1} ] _{:,i}\sim \mathcal{C} \mathcal{N} ( \mathbf{0},\sigma ^2\mathbf{I}_{N_{\mathrm{bs}}} )$ and $\tilde{\mathbf{n}}_{l,1}$ in \eqref{p_l_1} can be represented as
\begin{align}
	\tilde{\mathbf{n}}_{l,1}=[ \tilde{\mathbf{N}}_{1,1} ] _{:,l}=( \sqrt{Q_{\mathrm{rf},1}}N_{\mathrm{bs}} ) ^{-1}\mathbf{N}_{1,1}^{\mathrm{H}}[\hat{\mathbf{A}}_{N_{\mathrm{bs}}}] _{:,l}.
\end{align}

Define the covariance of $\tilde{\mathbf{n}}_{l,1}$ as $\mathbf{C}_{\tilde{\mathbf{n}}_{l,1}}=\mathbb{E} [ \tilde{\mathbf{n}}_{l,1}\tilde{\mathbf{n}}_{l,1}^{\mathrm{H}} ]$, whose elements are given by 
\begin{align}\label{C_n_l_1}
	&[\mathbf{C}_{\tilde{\mathbf{n}}_{l,1}}]_{i,j}=\mathbb{E} [[\tilde{\mathbf{n}}_{l,1}]_i[\tilde{\mathbf{n}}_{l,1}^{\mathrm{H}}]_j]=\mathbb{E} [[\tilde{\mathbf{n}}_{l,1}^{\mathrm{H}}]_j[\tilde{\mathbf{n}}_{l,1}]_i]\nonumber
	\\
	&=P_{1}^{-1}N_{\mathrm{bs}}^{-2}[\hat{\mathbf{A}}_{N_{\mathrm{bs}}}^{\mathrm{H}}]_{l,:}\mathbb{E} [[\mathbf{N}_{1,1}]_{:,j}[ \mathbf{N}_{1,1}^{\mathrm{H}} ] _{i,:}][\hat{\mathbf{A}}_{N_{\mathrm{bs}}}]_{:,l}.
\end{align}
For $\forall i \ne j$, $\mathbb{E} [[\mathbf{N}_{1,1}]_{:,j}[ \mathbf{N}_{1,1}^{\mathrm{H}} ] _{i,:}]=\mathbf{0}_{N_{\mathrm{bs}}\times N_{\mathrm{bs}}}$ and $[ \mathbf{C}_{\tilde{\mathbf{n}}_{l,1}} ] _{i,j}=0$.
For $\forall i=j$, $\mathbb{E} [[\mathbf{N}_{1,1}]_{:,j}[ \mathbf{N}_{1,1}^{\mathrm{H}} ] _{i,:}]=\sigma^2\mathbf{I}_{N_\mathrm{bs}}$ and $[ \mathbf{C}_{\tilde{\mathbf{n}}_{l,1}} ] _{i,i}$ can be further written as
\begin{align}
	&[\mathbf{C}_{\tilde{\mathbf{n}}_{l,1}}]_{i,i}=P_{1}^{-1}N_{\mathrm{bs}}^{-2}\sigma ^2[\hat{\mathbf{A}}_{N_{\mathrm{bs}}}^{\mathrm{H}}]_{l,:}\mathbf{I}_{N_\mathrm{bs}}[\hat{\mathbf{A}}_{N_{\mathrm{bs}}}]_{:,l}\nonumber
	\\
	&=P_{1}^{-1}N_{\mathrm{bs}}^{-2}\sigma ^2[\hat{\mathbf{A}}_{N_{\mathrm{bs}}}^{\mathrm{H}}\hat{\mathbf{A}}_{N_{\mathrm{bs}}}]_{l,l}\nonumber
	\\
	&=P_{1}^{-1}N_{\mathrm{bs}}^{-1}\sigma ^2<P_{1}^{-1}\sigma ^2.
\end{align}

In Stage III, the hybrid combining matrix is set to $\mathbf{W}_{k,1}=\mathbf{W}_{A}$. The noise $[ \tilde{\mathbf{N}}_{k,1} ] _{:,l}$ in \eqref{tilde_N_k_1} can thus be rewritten as
\begin{align}\label{tilde_N_k_1_opt}
	[\tilde{\mathbf{N}}_{k,1}]_{:,l}=&\frac{1}{\sqrt{P_k}}(\mathbf{W}_{k,1}\hat{\mathbf{A}}_{\mathrm{bs}})^{\dagger}\mathbf{W}_{k,1}[ \mathbf{N}_{k,1} ] _{:,l}\nonumber
	\\
	=&\frac{1}{\sqrt{P_k}}N_{\mathrm{bs}}^{-1}[\begin{matrix}
		\mathbf{I}_L&		\mathbf{0}_{L\times ( N_{\mathrm{rf}}-L )}\\
	\end{matrix}]\mathbf{W}_{A}[ \mathbf{N}_{k,1} ] _{:,l}\nonumber
	\\
	=&\frac{1}{\sqrt{P_k}}N_{\mathrm{bs}}^{-1}\hat{\mathbf{A}}_{N_{\mathrm{bs}}}^{\mathrm{H}}[ \mathbf{N}_{k,1} ] _{:,l},
\end{align}
with covariance
\begin{align}
	\mathbf{C}_{[\tilde{\mathbf{N}}_{k,1}]_{:,l}}=&\mathbb{E} [[ \tilde{\mathbf{N}}_{k,1} ] _{:,l}[ [\tilde{\mathbf{N}}_{k,1}]_{:,l} ] ^{\mathrm{H}}]\nonumber
	\\
	=&P_{k}^{-1}N_{\mathrm{bs}}^{-2}\hat{\mathbf{A}}_{N_{\mathrm{bs}}}^{\mathrm{H}}\mathbb{E} [[ \mathbf{N}_{k,1} ] _{:,l}[\mathbf{N}_{k,1}^{\mathrm{H}}]_{l,:}]\hat{\mathbf{A}}_{N_{\mathrm{bs}}}\nonumber
	\\
	=&P_{k}^{-1}N_{\mathrm{bs}}^{-2}\sigma ^2\hat{\mathbf{A}}_{N_{\mathrm{bs}}}^{\mathrm{H}}\hat{\mathbf{A}}_{N_{\mathrm{bs}}}\nonumber
	\\
	=&P_{k}^{-1}N_{\mathrm{bs}}^{-1}\sigma ^2\mathbf{I}_L<P_{k}^{-1}\sigma ^2\mathbf{I}_L.
\end{align}
Thus, we prove that the proposed design of the hybrid combining matrix $\mathbf{W}$ can maintain the noise power in \eqref{N_w1_1_1} and \eqref{tilde_N_k_1} at a relatively low level.

\subsubsection{Optimization of the RIS phase shift matrix $\mathbf{E}$}
The noise power in Sub-Stage II of Stage II is written as \eqref{n_w1_i_2}, which is a function of $\mathbf{E}$. Similar to Proposition 1, the noise power may be large if $\mathbf{E}$ is not properly chosen. To address this issue, we propose a novel choice for the RIS phase shift matrix:
\begin{align}\label{E_opt}
	\mathbf{E}_{A}=\hat{\mathbf{A}}_{\mathrm{RIS},r},
\end{align}
where $\hat{\mathbf{A}}_{\mathrm{RIS},r}$ is obtained in Stage I. Let $\mathbf{E}_{1,i}=\mathbf{E}_{A}$, so that $\mathbf{n}_{\beta ,i}$ in \eqref{n_w1_i_2} can be rewritten as
\begin{align}
	\mathbf{n}_{\beta ,i}=(\mathbf{E}_{1,i}^{\mathrm{H}}\hat{\mathbf{A}}_{\mathrm{RIS},r})^{\dagger}\tilde{\mathbf{n}}_{r,i}=M^{-1}\tilde{\mathbf{n}}_{r,i}.
\end{align}
The power of $\mathbf{n}_{\beta ,i}$ is $M^{-1}$ times that of $\tilde{\mathbf{n}}_{r,i}$.

\subsection{Pilot overhead analysis}
To simplify an analysis of the pilot overhead, we assume that all users have the same number of RF chains, antennas, propagation paths in their channels with the RIS, i.e., $J_1=\cdots =J_K=J$. In Stage I, the number of frames $V_0$ is a hyperparameter that satisfies $V_0\ge 1$. Thus the minimum pilot overhead in Stage I is $D$. According to \cite{1580791}, to find a $l$-sparse complex signal (vector) with dimension $n$, the number of measurements $m$ should satisfy $m \ge \mathcal{O} ( l\log ( n ) ) $. In Stage II, there exists a sparse recovery problem in Sub-Stage I. So the number of time slots $\tau _{1,1}$ is required to satisfy $\tau _{1,1}\ge \mathcal{O} ( J\log ( M ) ) $. In Sub-Stage II of Stage I, the number of time slots $\tau _{1,2}$ should satisfy $\tau _{1,2}\ge J$, otherwise the rank of the left inverse $(\mathbf{E}_{1,i}^{\mathrm{H}}\hat{\mathbf{A}}_{\mathrm{RIS},r})^{\dagger}$ is smaller than $J$. When the RIS phase shift matrix is optimized as $\mathbf{E}_{A}=\hat{\mathbf{A}}_{\mathrm{RIS},r}$, we have $\tau _{1,2}=J$. The number of antennas chosen to send pilots should satisfy $V_1\ge 1$. Thus, the minimum pilot overhead in Stage II is $\mathcal{O} ( J\log ( M ) ) +J$. In Stage III, there exists a sparse recovery problem in Sub-Stage I. Similar to Stage II, the number of time slots $\tau_{k,1}$ for user $k$ should satisfy $\tau _{k,1}\ge \mathcal{O} ( J\log ( M ) /L ) $. In Sub-Stage II, the number of frames $V_k$ must also satisfy $V_k\ge 1$, and $\tau _{k,2}\ge J/L$ is necessary to ensure that the left inverse $\mathbf{\Pi }^{\dagger}$ in \eqref{Pi_dagger} exists. Thus the minimum pilot overhead in Stage III is $\mathcal{O} ( J\log ( M ) /L ) +J/L$. Combining the above results, we see that the total minimum pilot overhead is $D+\mathcal{O} (J\log\mathrm{(}M))+J+( K-1 ) ( \mathcal{O} (J\log\mathrm{(}M)/L)+J/L ) $.

\begin{table*}[t]
	\centering
	\caption{Pilot Overhead of Different Channel Estimation Methods}
	\label{tab_pilot_overhead}
	\vspace{8pt}
	\resizebox{0.9\textwidth}{!}{
		\begin{tabular}{ccc}
			\toprule[1pt] 
			Architecture &Methods &Minimum Pilot Overhead \\
			\midrule[1pt]
			Fully-Digital   & Proposed algorithm &$\mathcal{O} (J\log\mathrm{(}M))+J+( K-1 ) ( \mathcal{O} (J\log\mathrm{(}M)/L)+J/L ) $ \\
			\midrule[0.5pt]
			Fully-Digital   & AoD-SOMP algorithm \cite{9919846} &$K\mathcal{O} ( J\log ( Q ) ) +J\mathcal{O} ( \log ( M ) ) +( K-1 )J \mathcal{O} ( \log ( M ) /L ) -JK$ \\
			\midrule[0.5pt]
			Hybrid   &Proposed algorithm &$D+\mathcal{O} (J\log\mathrm{(}M))+J+( K-1 ) ( \mathcal{O} (J\log\mathrm{(}M)/L)+J/L ) $ \\
			\midrule[0.5pt]
			Hybrid   &Extension of AoD-SOMP algorithm &$D(K\mathcal{O} ( J\log ( Q ) ) +J\mathcal{O} ( \log ( M ) ) +( K-1 )J \mathcal{O} ( \log ( M ) /L ) -JK)$ \\
			\bottomrule[1pt]
		\end{tabular}
	}
	\vspace{-0.2cm}
\end{table*}

\section{SIMULATION RESULTS}
In this section, simulation results are presented to evaluate the performance of the proposed three-stage estimation scheme. The performance metric is defined by the normalized mean square error (NMSE) of the cascaded channel matrix as $\mathbb{E} [ ( \sum_{k=1}^K{\| \hat{\mathbf{G}}_k-\mathbf{G}_k \| _{F}^{2}} ) /( \sum_{k=1}^K{\| \mathbf{G}_k \| _{F}^{2}} ) ]  $. All the results are obtained by averaging over 10000 channel realizations. The distance between the BS and the RIS is $d_{\mathrm{br}}=80$m and the distance between the RIS and users is $d_{\mathrm{ru}}=40$m. The channel gains are drawn from complex Gaussian distributions $\alpha _l\sim \mathcal{C} \mathcal{N} ( 0,10^{-3}d_{\mathrm{br}}^{-2.2} )$ and $\beta _{j_k}\sim \mathcal{C} \mathcal{N} ( 0,10^{-3}d_{\mathrm{ru}}^{-2.8} )$. The number of users is $K=4$ and all the users are equipped with the same number of antennas and RF chains, i.e., $Q_1=\cdots =Q_k=Q$ and $Q_{\mathrm{rf},1}=\cdots =Q_{\mathrm{rf},k}=Q_{\mathrm{rf}}$. The SNR is defined as $\mathrm{SNR}=10\log\mathrm{(}10^{-6}d_{\mathrm{br}}^{-2.2}d_{\mathrm{ru}}^{-2.8}P/\sigma ^2) $, and the transmit power of users is set to $P_1=\cdots =P_K=P$. The element spacings of the RIS, the BS and the user arrays are $d_{\mathrm{ris}}=d_{\mathrm{\mathrm{bs}}}=d_{\mathrm{\mathrm{ue}}}=\frac{\lambda}{2}$. The number of elements in the RIS is $M=256$ with $M_1=M_2=16$, and the number of antennas at the BS is $N_{\mathrm{bs}}=128$. The number of paths from the RIS to the BS is $L=4$ and the number of paths from users to the RIS is $J_1=\cdots=J_K=J=4$. We compare the performance of the following channel estimation algorithms:
\begin{figure}[t]
	\centering
	\includegraphics[width=0.4\textwidth]{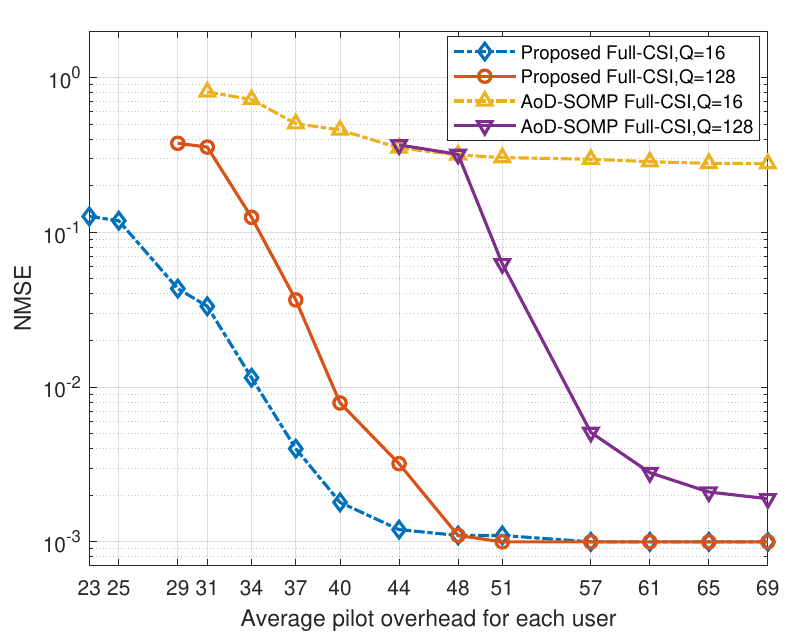}
	\caption{NMSE versus average pilot overhead $T$ of each user with SNR = 5 dB and a fully-digital architecture.}
	\label{simul5}
	\vspace{-0.5cm}
\end{figure}

\begin{itemize}
	\item Proposed Full-CSI: The cascaded channels for all users are estimated using the proposed three-stage estimation strategy, and the OMP algorithm is used to solve the sparsity recovery problem. 
        \item Proposed Oracle: In this approach, the BS is assumed to 
    have perfect angle information so that we can directly proceed to Stage II and III. The channel gains are estimated using LS. The performance of this method is an upper bound for the Proposed Full-CSI method.
	\item Ext AoD-SOMP Full-CSI\cite{9919846}: In this approach, the AoDs of the users are first estimated using the SOMP algorithm. Then the MIMO cascaded channel is found by estimating the individual MISO cascaded channels. We extend the approach in \cite{9919846} to hybrid communication systems based on \cite{8493600} by repeatedly sending pilot signals to reconstruct the uncompressed received signals.
        \item Ext AoD-SOMP Oracle: This method also assumes that the BS 
    has perfect angle information, but otherwise follows the estimation approach in \cite{9919846}. This algorithm can be regarded as the performance upper bound of the Ext AoD-SOMP Full-CSI method and as a benchmark for the Proposed Oracle to demonstrate the accuracy of the gain estimation for the proposed method.
\end{itemize}

As shown in Table \ref{tab_pilot_overhead}, there is a significant difference in pilot overhead between the above methods for hybrid architectures, so we first analyze the impact of the average pilot overhead $T$ on the estimation performance for a fully-digital architecture in Fig. \ref{simul5} when the SNR is 5dB. As expected, more pilots lead to better NMSE performance for all methods. As shown in Fig. \ref{simul5}, the proposed method has a lower NMSE than the existing ``AoD-SOMP" benchmark of \cite{9919846} with much less pilot overhead for all tested values of $T$ and $Q$. For the proposed method,  more pilots are required with  $Q=128$ antennas to achieve the same NMSE as with $Q=16$ antennas. These observations can be explained as follows. AoD-SOMP estimates both gains and angles even though only angle estimates are needed, which results in a waste of pilots and poor estimation performance when only few pilots are available. In addition, due to the strong correlation between atoms in the overcomplete dictionary when the number of antennas is small, the estimation performance when $Q$ = 16 is poor. For the proposed method, a larger number of antennas obviously leads to larger-scale channels that require more pilots to estimate. Moreover, the AoDs at the users are estimated by solving a one-dimensional search problem without estimating unnecessary parameters, which significantly reduces the required pilot overhead and avoids restrictions on the number of antennas.

\begin{figure}[t]
	\centering
	\includegraphics[width=0.4\textwidth]{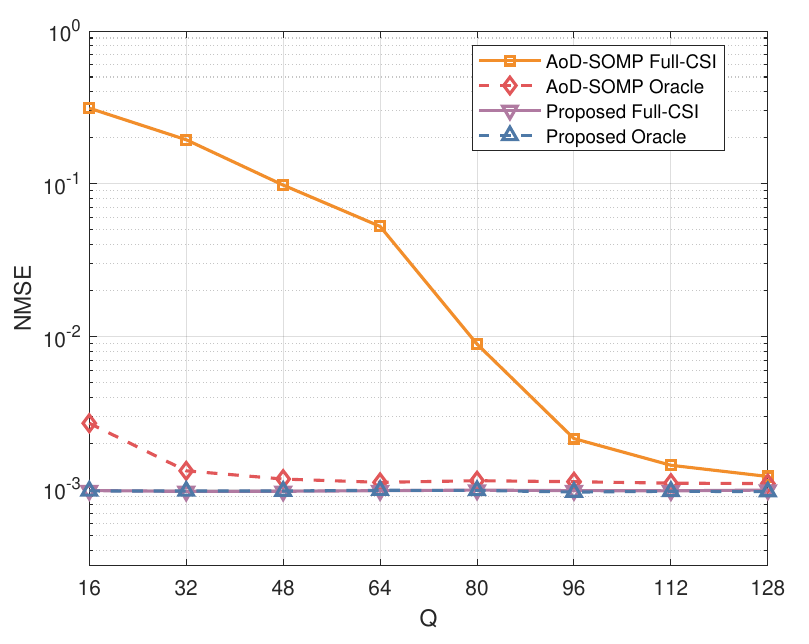}
	\caption{NMSE versus number of user antennas $Q$ for fully-digital architecture when SNR = 5dB.}
	\label{simul2}
	\vspace{-0.5cm}
\end{figure}

Fig. \ref{simul2} illustrates the impact of the number of user antennas when the SNR is 5dB for a fully-digital architecture. In order to eliminate the influence of the
number of pilots, we use the same pilot overhead and choose it to be sufficiently large for all methods. We see that the NMSE of the AoD-SOMP Oracle is essentially the same as for the Proposed Oracle, which indicates that the performance bound of these two methods is similar. However, for AoD-SOMP there is an NMSE difference of approximately two orders of magnitude between the Oracle and the Full-CSI case when $Q \leq 48$, which indicates poor angle estimation accuracy. When $64\leq Q \leq 96$, the estimation performance of the Full-CSI method improves significantly, and achieves nearly the same performance as the Oracle when $Q \geq 112$. As before, the poor performance for small $Q$ is due to the strong correlation between atoms in the dictionary for the SOMP Algorithm. By contrast, for the proposed method, the performance of Full-CSI is almost identical to that of Oracle regardless of the number of antennas, which indicates good angle estimation accuracy and performance that is not limited by the number of user antennas.

\begin{figure}[t]
	\centering
	\includegraphics[width=0.4\textwidth]{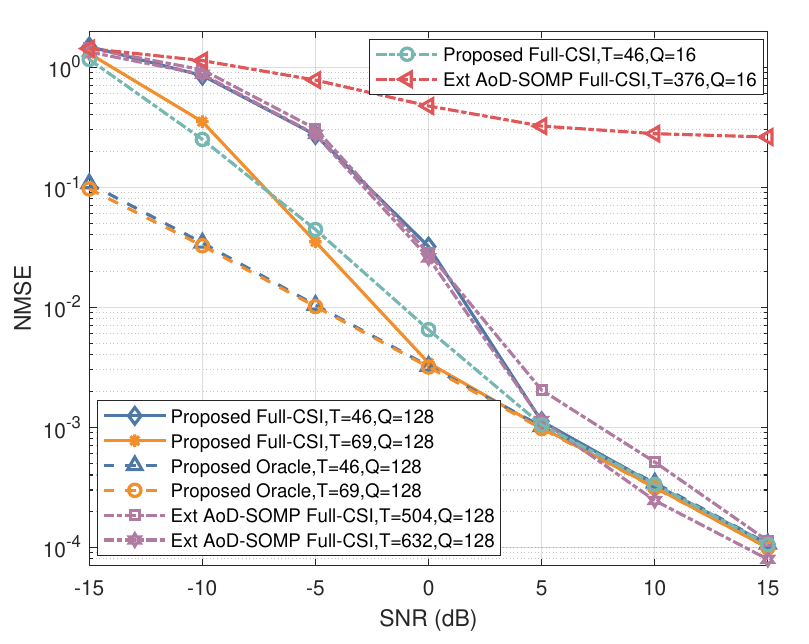}
	\caption{NMSE versus SNR for the hybrid architecture with $N_{\mathrm{rf}}=16$, $Q_{\mathrm{rf}}=2$.}
	\label{simul1}
	\vspace{-0.5cm}
\end{figure}

Fig. \ref{simul1} shows the NMSE performance as a function of SNR for the considered methods for the hybrid architecture. Due to the need to reconstruct the uncompressed received signals when directly extending the existing SOMP-based method to the hybrid case, the pilot overhead of Ext AoD-SOMP is large ($T$ = 376) compared to the proposed method ($T$=46). When $Q=16$, the NMSE of Ext AoD-SOMP Full-CSI is poor even when the SNR is high, while the NMSE rapidly decreases with SNR for the Proposed Full-CSI method. When $Q=128$, NMSE of all methods improves with SNR. However, at low SNR (SNR $\le$ 5 dB), there exists a performance gap between the Full-CSI and Oracle approach resulting from estimation errors in both the gain and angle domains. For the Proposed Full-CSI method, the performance gap can be alleviated by increasing the average pilot overhead $T$ from 46 to 69. By contrast, the NMSE of Ext AoD-SOMP Full-CSI still has performance issues even with a high pilot overhead. 

\begin{figure}[t]
	\centering
	\includegraphics[width=0.4\textwidth]{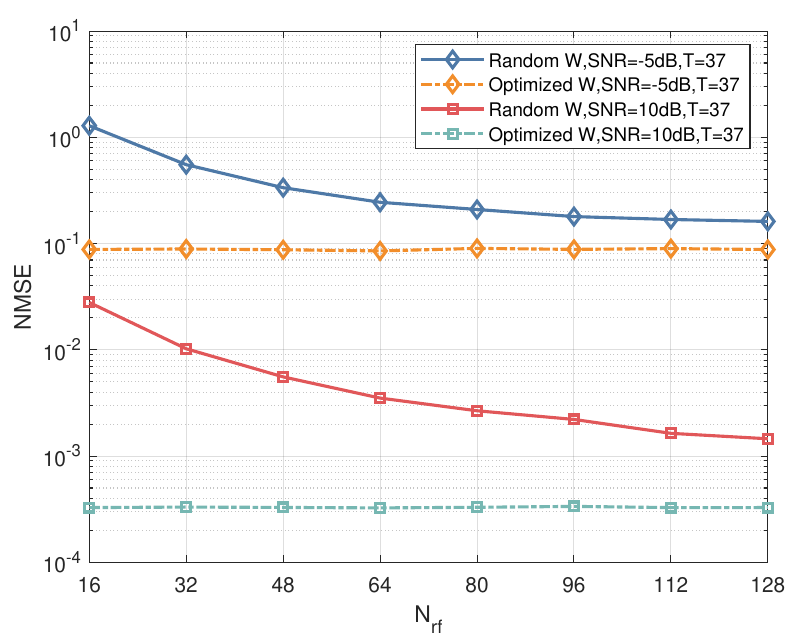}
	\caption{Performance gain achieved by optimizing the hybrid combining matrix versus the number of RF chains at the BS, when $Q_{\mathrm{rf}}=2$, $Q=16$.}
	\label{simul3}
	\vspace{-0.4cm}
\end{figure}

In Fig. \ref{simul3}, we explore the performance gain achieved by optimizing the hybrid combining matrix $\mathbf{W}$ at the BS. The label ``Random W" indicates that the elements of the hybrid combining matrix are given by $[ \mathbf{W} ] _{m,n}=\exp ( i2\pi a_{m,n})$, where $a_{m,n}$ is an independently and identically distributed uniform random variable, i.e., $a_{m,n}\sim \mathcal{U} ( 0,1 ) $. ``Optimized W" means that we construct the hybrid combing matrix as $\mathbf{W}_{A}$ in \eqref{W_opt}. The performance of Random W gradually approaches that for Optimized W as the number of RF chains increases, possibly because a larger number of RF chains increases the dimension of the hybrid combining matrix, which increases the modulus of the elements in $\mathbf{\Lambda }$ in \eqref{SVD_W_1_1} and \eqref{SVD_W_k_1}. However, $\mathbf{W}_{1,1}\mathbf{W}_{1,1}^{\mathrm{H}}$ and $\mathbf{W}_{k,1}\mathbf{W}_{k,1}^{\mathrm{H}}$ in \eqref{C_n_l_1_appendix} and \eqref{C_n_wk_1_1_appendix} remain relatively stable as the number of RF chains increases, which means the modulus of the elements on the diagonal remains unchanged and is far larger than that of other elements. Thus, 
combining the above two observations, there is a higher probability that the noise power in \eqref{C_n_l_1_appendix} and \eqref{C_n_wk_1_1_appendix} will decrease. 

\begin{figure}[t]
	\centering
	\includegraphics[width=0.4\textwidth]{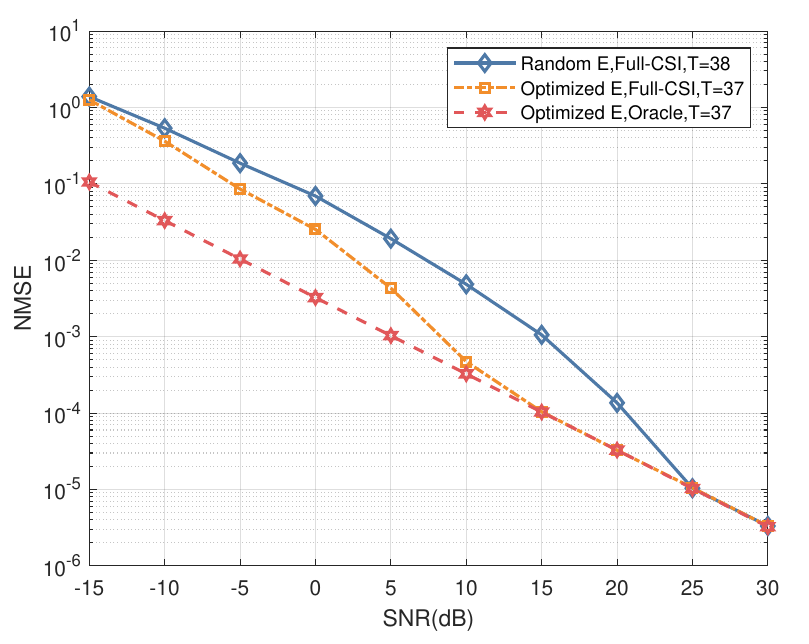}
	\caption{Performance achieved by optimizing the RIS phase shift matrix versus SNR, when $Q_{\mathrm{rf}}=2$, $Q=16$ and $N_{\mathrm{rf}}=16$.}
	\label{simul4}
	\vspace{-0.4cm}
\end{figure}

The benefit of optimizing of the RIS phase shift matrix $\mathbf{E}$ is illustrated in Fig. \ref{simul4}. ``Random E" indicates that the elements of the RIS phase shift matrix are given by $[ \mathbf{E} ] _{m,n}=\exp ( i2\pi a_{m,n})$, where as before, $a_{m,n}\sim \mathcal{U} ( 0,1 ) $. ``Optimized E" means that we construct the RIS phase shift matrix as $\mathbf{E}_{A}$ in \eqref{E_opt}. As shown in Fig. \ref{simul4}, the performance gap between Optimized E and Random E for the Full-CSI methods first increases and then decreases when $-15$dB $\le \mathrm{SNR}\le 20$dB, and then becomes marginal when $\mathrm{SNR}\ge 25$dB. Furthermore, the performance of Optimized E reaches that of the Oracle more quickly, i.e., when $\mathrm{SNR}=10$dB. These results illustrate that optimization of the RIS phase shifts leads to better estimation accuracy for typical SNRs in the range $-15$dB $\le \mathrm{SNR}\le 25$dB), and the proposed method can reach the theoretical performance upper bound at a relatively low SNR due to the significant reduction in noise power. 

\section{CONCLUSIONS}
In this paper, we proposed a novel three-stage cascaded channel estimation scheme for an RIS-aided uplink MU-MIMO mmWave system assuming a hybrid RF architecture. By exploiting the fact that only estimates of the common AoAs are needed to reconstruct the uncompressed received signals, we separated the estimation of AoAs and other parameters into two independent stages, leading to a significant reduction in pilot overhead. In addition, our method not only exploits the inherent relationship between the gains of the cascaded subchannels, but also estimates the AoDs at the users by means of a simple one-dimensional search, which leads to a significant improvement in estimation accuracy when the number of antennas is small. Moreover,  optimization of the hybrid combining and precoding matrices and the RIS phase shifts was also explored to reduce the noise power. Simulation results showed that the proposed method outperforms existing SOMP-based methods with much less pilot overhead.

\begin{appendices}\label{proposition_1}
	\section{Proof of Proposition 1}
	\subsection{Stage II noise in \eqref{p_l_1}}
		The SVD of $\mathbf{W}_{1,1}\hat{\mathbf{A}}_{N_\mathrm{bs}}$ in \eqref{N_w1_1_1} can be represented as
	\begin{align}\label{SVD_W_1_1}
		\mathbf{W}_{1,1}\hat{\mathbf{A}}_{N_\mathrm{bs}}=\mathbf{U}[ \begin{matrix}
			\mathbf{\Lambda }&		\mathbf{0}_{L\times \left( N_{\mathrm{rf}}-L \right)}\\
		\end{matrix} ] ^{\mathrm{T}}\mathbf{V}^{\mathrm{H}},
	\end{align}
	where both $\mathbf{U}\in \mathbb{C} ^{N_{\mathrm{rf}}\times N_{\mathrm{rf}}}$ and $\mathbf{V}\in \mathbb{C} ^{L\times L}$ are unitary matrices, and $\mathbf{\Lambda }\in \mathbb{C} ^{L\times L}$ is a diagonal matrix. If $(\mathbf{W}_{1,1}\hat{\mathbf{A}}_{N_{\mathrm{bs}}})^{\dagger}$ exists, we can derive 
	\begin{align}
		(\mathbf{W}_{1,1}\hat{\mathbf{A}}_{N_{\mathrm{bs}}})^{\dagger}=&((\mathbf{W}_{1,1}\hat{\mathbf{A}}_{N_{\mathrm{bs}}})^{\mathrm{H}}(\mathbf{W}_{1,1}\hat{\mathbf{A}}_{N_{\mathrm{bs}}}))^{-1}(\mathbf{W}_{1,1}\hat{\mathbf{A}}_{N_{\mathrm{bs}}})^{\mathrm{H}}\nonumber
		\\
		=&\mathbf{V}\left( \mathbf{\Lambda }^{\mathrm{H}}\mathbf{\Lambda } \right) ^{-1}\mathbf{V}^{\mathrm{H}}\mathbf{V}\left[ \begin{matrix}
			\mathbf{\Lambda }^{\mathrm{H}}&		\mathbf{0}_{L\times \left( N_{\mathrm{rf}}-L \right)}\\
		\end{matrix} \right] \mathbf{U}^{\mathrm{H}}\nonumber
		\\
		=&\mathbf{V}\left( \mathbf{\Lambda }^{\mathrm{H}}\mathbf{\Lambda } \right) ^{-1}\left[ \begin{matrix}
			\mathbf{\Lambda }^{\mathrm{H}}&		\mathbf{0}_{L\times \left( N_{\mathrm{rf}}-L \right)}\\
		\end{matrix} \right] \mathbf{U}^{\mathrm{H}}.
	\end{align}
	The noise power in \eqref{N_w1_1_1} can be rewritten as
	\begin{align}
		&\tilde{\mathbf{N}}_{1,1}=\frac{1}{\sqrt{P_1}}[(\mathbf{W}_{1,1}\hat{\mathbf{A}}_{N_{\mathrm{bs}}})^{\dagger}\mathbf{W}_{1,1}\mathbf{N}_{1,1}]^{\mathrm{H}}\nonumber
		\\
		&=\frac{1}{\sqrt{P_1}}[\mathbf{V}\left( \mathbf{\Lambda }^{\mathrm{H}}\mathbf{\Lambda } \right) ^{-1}\left[ \begin{matrix}
			\mathbf{\Lambda }^{\mathrm{H}}&		\mathbf{0}_{L\times \left( N_{\mathrm{rf}}-L \right)}\\
		\end{matrix} \right] \mathbf{U}^{\mathrm{H}}\mathbf{W}_{1,1}\mathbf{N}_{1,1}]^{\mathrm{H}}\nonumber
		\\
		&=\frac{1}{\sqrt{P_1}}\mathbf{N}_{1,1}^{\mathrm{H}}\mathbf{W}_{1,1}^{\mathrm{H}}\mathbf{U}\left[ \begin{matrix}
			\mathbf{\Lambda }&		\mathbf{0}_{L\times \left( N_{\mathrm{rf}}-L \right)}\\
		\end{matrix} \right] ^{\mathrm{T}}\left( \mathbf{\Lambda }^{\mathrm{H}}\mathbf{\Lambda } \right) ^{-1}\mathbf{V}^{\mathrm{H}}.
	\end{align}
	Then, $\tilde{\mathbf{n}}_{l,1}$ in \eqref{p_l_1} can be rewritten as
	\begin{align}
		&\tilde{\mathbf{n}}_{l,1}=[\tilde{\mathbf{N}}_{1,1}]_{:,l}\nonumber
		\\
		&=\frac{1}{\sqrt{P_1}}\mathbf{N}_{1,1}^{\mathrm{H}}\mathbf{W}_{1,1}^{\mathrm{H}}\mathbf{U}\left[ \begin{array}{c}
			\mathbf{\Lambda }\left( \mathbf{\Lambda }^{\mathrm{H}}\mathbf{\Lambda } \right) ^{-1}\\
			\mathbf{0}_{\left( N_\mathrm{rf}-L \right) \times L}\\
		\end{array} \right] [\mathbf{V}^{\mathrm{H}}]_{:,l},
	\end{align}
	with covariance 
	\begin{align}
		&[\mathbf{C}_{\tilde{\mathbf{n}}_{l,1}}]_{i,j}=\mathbb{E} [[\tilde{\mathbf{n}}_{l,1}]_{j}^{\mathrm{H}}[\tilde{\mathbf{n}}_{l,1}]_i]\nonumber
		\\
		&=P_{1}^{-1}[\mathbf{V}]_{l,:}\left[ \begin{array}{c}
			\mathbf{\Lambda }\left( \mathbf{\Lambda }^{\mathrm{H}}\mathbf{\Lambda } \right) ^{-1}\\
			\mathbf{0}_{\left( N_\mathrm{rf}-L \right) \times L}\\
		\end{array} \right] ^{\mathrm{H}}\mathbf{U}^{\mathrm{H}}\mathbf{W}_{1,1}
		\\
		&\times\mathbb{E} [[\mathbf{N}_{1,1}]_{:,j}\left[ \mathbf{N}_{1,1}^{\mathrm{H}} \right] _{i,:}]\mathbf{W}_{1,1}^{\mathrm{H}}\mathbf{U}\left[ \begin{array}{c}
			\mathbf{\Lambda }\left( \mathbf{\Lambda }^{\mathrm{H}}\mathbf{\Lambda } \right) ^{-1}\\
			\mathbf{0}_{\left( N_\mathrm{rf}-L \right) \times L}\\
		\end{array} \right] [\mathbf{V}^{\mathrm{H}}]_{:,l}.\nonumber
	\end{align}
    
	When $i \ne j$, $\mathbb{E} [[\mathbf{N}_{1,1}]_{:,j}[ \mathbf{N}_{1,1}^{\mathrm{H}} ] _{i,:}]=\mathbf{0}_{N_{\mathrm{bs}}\times N_{\mathrm{bs}}}$, thus $[ \mathbf{C}_{\tilde{\mathbf{n}}_{l,1}} ] _{i,j}=0$. When $i=j$, $\mathbb{E} [[\mathbf{N}_{1,1}]_{:,j}[ \mathbf{N}_{1,1}^{\mathrm{H}} ] _{i,:}]=\sigma^2\mathbf{I}_{N_{\mathrm{bs}}}$, and hence
	\begin{align}\label{C_n_l_1_appendix}
		[\mathbf{C}_{\tilde{\mathbf{n}}_{l,1}}]_{i,i}=&P_{1}^{-1}\sigma ^2[\mathbf{V}]_{l,:}\left[ \begin{array}{c}
			\mathbf{\Lambda }\left( \mathbf{\Lambda }^{\mathrm{H}}\mathbf{\Lambda } \right) ^{-1}\\
			\mathbf{0}_{\left( N_\mathrm{rf}-L \right) \times L}\\
		\end{array} \right] ^{\mathrm{H}}\mathbf{U}^{\mathrm{H}}\mathbf{W}_{1,1}\nonumber
		\\
		&\times\mathbf{W}_{1,1}^{\mathrm{H}}\mathbf{U}\left[ \begin{array}{c}
			\mathbf{\Lambda }\left( \mathbf{\Lambda }^{\mathrm{H}}\mathbf{\Lambda } \right) ^{-1}\\
			\mathbf{0}_{\left( N_\mathrm{rf}-L \right) \times L}\\
		\end{array} \right] [\mathbf{V}^{\mathrm{H}}]_{:,l}.
	\end{align} 
	Define $\lambda _{\max}$ as the element of $\mathrm{diag}\{\mathbf{\Lambda }^{\mathrm{H}}\mathbf{\Lambda }\} $ the largest value. When the hybrid combining matrix $\mathbf{W}_{1,1}$ satisfies the following condition
	\begin{align}
		\mathbf{W}_{1,1}\mathbf{W}_{1,1}^{\mathrm{H}}\succeq \lambda _{\max}\mathbf{I}_L,
	\end{align} 
	\eqref{C_n_l_1_appendix} can be scaled as follows
	\begin{align}
		[ \mathbf{C}_{\tilde{\mathbf{n}}_{l,1}} ] _{i,i}\geq& P_{1}^{-1}\sigma ^2\lambda _{\max}[\mathbf{V}]_{l,:}( \mathbf{\Lambda }^{\mathrm{H}}\mathbf{\Lambda } ) ^{-1}[ \mathbf{V}^{\mathrm{H}} ] _{:,l}\nonumber
		\\
		\geq& P_{1}^{-1}\sigma ^2[\mathbf{V}]_{l,:}\mathbf{I}_L[ \mathbf{V}^{\mathrm{H}} ] _{:,l}\nonumber
		\\
		=&P_{1}^{-1}\sigma ^2.
	\end{align}
	
	\subsection{Stage III noise in \eqref{tilde_N_k_1}}
	As in the first section of the appendix, $\mathbf{W}_{k,1}\hat{\mathbf{A}}_{\mathrm{bs}}$ in \eqref{tilde_N_k_1} can be decomposed as
	\begin{align}\label{SVD_W_k_1}
		\mathbf{W}_{k,1}\hat{\mathbf{A}}_{N_{bs}}=\mathbf{U}[ \begin{matrix}
			\mathbf{\Lambda }&		\mathbf{0}_{L\times \left( N_\mathrm{rf}-L \right)}\\
		\end{matrix} ] ^{\mathrm{T}}\mathbf{V}^{\mathrm{H}}, 
	\end{align}
	and the noise in \eqref{tilde_N_k_1} can be rewritten as
	\begin{align}
		[ \tilde{\mathbf{N}}_{k,1} ] _{:,l}=\frac{1}{\sqrt{P_k}}\mathbf{V}\left[ \begin{array}{c}
			\mathbf{\Lambda }( \mathbf{\Lambda }^{\mathrm{H}}\mathbf{\Lambda } ) ^{-1}\\
			\mathbf{0}_{\left( N_\mathrm{rf}-L \right) \times L}\\
		\end{array} \right] ^{\mathrm{H}}\mathbf{U}^{\mathrm{H}}\mathbf{W}_{k,1}[ \mathbf{N}_{k,1} ] _{:,l}
	\end{align}
	with covariance
	\begin{align}\label{C_n_wk_1_1_appendix}
		&\left[\mathbf{C}_{[ \tilde{\mathbf{N}}_{k,1} ] _{:,l}} \right] _{i,i}=\mathbb{E} [[ \tilde{\mathbf{N}}_{k,1} ] _{i,l}[ \tilde{\mathbf{N}}_{k,1} ] _{i,l}^{\mathrm{H}}]\nonumber
		\\
		&=P_{k}^{-1}[ \mathbf{V} ] _{i,:}\left[ \begin{array}{c}
			\mathbf{\Lambda }( \mathbf{\Lambda }^{\mathrm{H}}\mathbf{\Lambda } ) ^{-1}\\
			\mathbf{0}_{( N_\mathrm{rf}-L ) \times L}\\
		\end{array} \right] ^{\mathrm{H}}\mathbf{U}^{\mathrm{H}}\mathbf{W}_{k,1}\mathbf{W}_{k,1}^{\mathrm{H}}\nonumber
		\\
		&\times \mathbb{E} [[ \mathbf{N}_{k,1} ] _{:,l}[\mathbf{N}_{k,1}^{\mathrm{H}}]_{l,:}]\mathbf{U}\left[ \begin{array}{c}
			\mathbf{\Lambda }\left( \mathbf{\Lambda }^{\mathrm{H}}\mathbf{\Lambda } \right) ^{-1}\\
			\mathbf{0}_{( N_\mathrm{rf}-L ) \times L}\\
		\end{array} \right] [ \mathbf{V}^{\mathrm{H}} ] _{:,i}\nonumber
		\\
		&=P_{k}^{-1}\sigma ^2[\mathbf{V}]_{i,:}\left[ \begin{array}{c}
			\mathbf{\Lambda }(\mathbf{\Lambda }^{\mathrm{H}}\mathbf{\Lambda })^{-1}\\
			\mathbf{0}_{(N_\mathrm{rf}-L)\times L}\\
		\end{array} \right] ^{\mathrm{H}}\mathbf{U}^{\mathrm{H}}\nonumber
		\\
		&\times \mathbf{W}_{k,1}\mathbf{W}_{k,1}^{\mathrm{H}}\mathbf{U}\left[ \begin{array}{c}
			\mathbf{\Lambda }\left( \mathbf{\Lambda }^{\mathrm{H}}\mathbf{\Lambda } \right) ^{-1}\\
			\mathbf{0}_{(N_\mathrm{rf}-L)\times L}\\
		\end{array} \right] [\mathbf{V}^{\mathrm{H}}]_{:,i}.
	\end{align}
	When the hybrid combining matrix $\mathbf{W}_{k,1}$ satisfies the following condition
	\begin{align}
		\mathbf{W}_{k,1}\mathbf{W}_{k,1}^{\mathrm{H}}\succeq \lambda _{\max}\mathbf{I}_L,
	\end{align} 
	\eqref{C_n_wk_1_1_appendix} be scaled as follows
	\begin{align}
		[\mathbf{C}_{[ \tilde{\mathbf{N}}_{k,1} ] _{:,l}} ] _{i,i}\geq& P_{k}^{-1}\sigma ^2\lambda _{\max}[ \mathbf{V} ] _{i,:}\left( \mathbf{\Lambda }^{\mathrm{H}}\mathbf{\Lambda } \right) ^{-1}[ \mathbf{V}^{\mathrm{H}} ] _{:,i}\nonumber
		\\
		\geq& P_{k}^{-1}\sigma ^2.
	\end{align}
	This completes the proof.

\end{appendices}

\bibliographystyle{IEEEtran}
\bibliography{IEEEabrv,paper}

\end{document}